\documentclass[usenatbib]{mnras}
\usepackage[utf8]{inputenc}
\usepackage{aas_macros}
\usepackage{amsfonts}
\usepackage{amsmath}
\usepackage{amssymb}
\usepackage{graphicx}
\usepackage[dvipsnames]{xcolor}
\usepackage{hyperref}
\hypersetup{colorlinks=true,allcolors=teal}
\usepackage{microtype}

\newcommand{\Hi}{\textsc{Hi}}
\newcommand{\mHi}{\ensuremath{m_{\Hi}}}

\newcommand{\p}{\ensuremath{\partial}}

\newcommand{\Msun}{\ensuremath{M_{\odot}}}
\newcommand{\Mh}{\ensuremath{h^{-1}M_{\odot}}}
\newcommand{\Mhsq}{\ensuremath{h^{-2}M_{\odot}}}
\newcommand{\Mpch}{\ensuremath{h^{-1}{\rm Mpc}}}
\newcommand{\kpch}{\ensuremath{h^{-1}{\rm kpc}}}

\newcommand{\avg}[1]{\ensuremath{\left\langle \,#1\, \right\rangle}}
\newcommand{\e}[1]{\ensuremath{{\rm e}^{#1}}}

\newcommand{\der}{\ensuremath{{\rm d}}}

\newcommand{\erfc}[1]{\ensuremath{{\rm erfc}\left(#1\right)}}
\newcommand{\erf}[1]{\ensuremath{{\rm erf}\left(#1\right)}}

\newcommand{\eqn}[1]{equation~\eqref{#1}}
\newcommand{\eqns}[1]{equations~\eqref{#1}}

\newcommand{\be}{\begin{equation}}
\newcommand{\ee}{\end{equation}}
\newcommand{\Cal}[1]{\ensuremath{\mathcal{#1}}}

\title[Colourful mocks]{Multi-wavelength mock galaxy catalogs of the low-redshift Universe} 
\author[Paranjape et al.]
{
{\parbox[t]{\textwidth}{
Aseem Paranjape$^{1}$\thanks{E-mail: aseem@iucaa.in},
Tirthankar Roy Choudhury$^{2}$\thanks{E-mail: tirth@ncra.tifr.res.in} \&
Ravi K. Sheth$^{3,4}$\thanks{E-mail: shethrk@physics.upenn.edu}
}}
\\\, \\
 $^1$ Inter-University Centre for Astronomy \& Astrophysics,
      Ganeshkhind, Post Bag 4, Pune 411007, India\\
  $^2$ National Centre for Radio Astrophysics, TIFR, Post Bag 3, Ganeshkhind, Pune 411007, India\\
  $^3$ Center for Particle Cosmology, University of Pennsylvania, 209 S. 33rd St., Philadelphia, PA 19104, USA\\
 $^4$ The Abdus Salam International Center for Theoretical Physics, Strada Costiera, 11, Trieste 34151, Italy}

\begin{document}
\label{firstpage}
\pagerange{\pageref{firstpage}--\pageref{lastpage}}
\maketitle

\begin{abstract}
We present a new suite of mock galaxy catalogs mimicking the low-redshift Universe, based on an updated halo occupation distribution (HOD) model and a scaling relation between optical properties and the neutral hydrogen (\Hi) content of galaxies. 
Our algorithm is constrained by observations of the luminosity function and luminosity- and colour-dependent clustering of SDSS galaxies, as well as the \Hi\ mass function and \Hi-dependent clustering of massive \Hi-selected galaxies in the ALFALFA survey. 
Mock central and satellite galaxies with realistic values of $r$-band luminosity, $g-r$ and $u-r$ colour, stellar mass and \Hi\ mass are populated in an $N$-body simulation, inheriting a number of properties of the density and tidal environment of their host halos. 
The host halo of each central galaxy is also `baryonified' with realistic spatial distributions of stars as well as hot and cold gas,  along with the corresponding rotation curve. 
Our default HOD assumes that galaxy properties are a function of group halo mass alone, and can optionally include effects such as galactic conformity and colour-dependent galaxy assembly bias. 
The mocks predict the relation between the stellar mass and \Hi\ mass of massive \Hi\ galaxies, as well as the 2-point cross-correlation function of spatially co-located optical and \Hi-selected samples. 
They enable novel null tests for galaxy assembly bias, provide predictions for the \Hi\ velocity width function, and clarify the origin and universality of the radial acceleration relation in the $\Lambda$CDM framework. 
\end{abstract}


\section{Introduction}
\label{sec:intro}
Contemporary studies of galaxy evolution and cosmology must explore a multitude of physical and statistical properties of the observed large-scale structure of the Universe.  The computational and technical challenges involved in any such analysis, whether observational or theoretical, mean that mock galaxy catalogs are now a staple tool of cosmological analyses.

Current and upcoming large-volume surveys of the Universe, aiming to extract cosmological information using observables including the redshift space clustering of galaxies at large and small scales, weak lensing, the abundances of clusters and voids, etc., increasingly rely on the use of mock galaxy catalogs for a variety of applications. 
Apart from calibrating expected measurement covariances and end-to-end pipeline testing \citep{mao+18-descqa}, such catalogs also serve as excellent test-beds for exploring ideas related to the galaxy-dark matter connection, the nature of dark matter or the predictions of alternative gravity theories, and the effects of dark energy. As such, it is critical to develop and calibrate mock-making algorithms, constrained by existing observations, that can accurately account for the multi-scale, multi-probe connection between baryonic matter in and around galaxies and the dark cosmic web in which these galaxies reside. This is the primary motivation behind the present work.

Computationally speaking, the most efficient algorithms are those which model the baryon-dark matter connection using empirical, statistical tools that are motivated by the Halo Model \cite[see][for a review]{cs02}.  These include the halo occupation distribution \cite[HOD;][]{zehavi+11}, conditional luminosity function \cite[CLF;][and references therein]{elucidV} or subhalo abundance matching \cite[SHAM;][and references therein]{bwhc19} prescriptions that are constrained by observed galaxy abundances and clustering. At the other end of the spectrum, lie full-fledged cosmological hydrodynamical simulations of galaxy formation \citep[e.g.,][]{vogelsberger+14,dubois+14,schaye+15,springel+18}, arguably the most realistic and most expensive tool in computational cosmology. 
Semi-analytical models (SAMs), which evolve simplified physical descriptions of  galaxy formation and evolution within the cosmic web of gravity-only simulations, lie somewhere between full simulations and empirical models, in terms of both computational complexity as well as fidelity to observational constraints \cite[for a recent review, see][]{sd15}. The present work focuses on HOD models.

Mock-making algorithms based on the HOD or SHAM frameworks have been frequently used in the literature in conjunction with large-volume surveys at low and intermediate redshifts \citep{Manera13,dtp13,delatorre+13,kitaura+16,mao+18-descqa,alam+20,cheng+20,sugiyama+20}. These algorithms typically segregate into those describing stellar populations and overall star formation activity (constrained by a multitude of galaxy surveys at low- and high-redshift spanning wavelengths from the infrared to optical to ultraviolet), and others focused on the distribution of gas, primarily in the form of neutral hydrogen (\Hi, constrained by radio wavelength observations, typically at low-redshift). We broadly refer to the former category using the label `optical' and the latter as `\Hi' algorithms. All such algorithms typically rely on dark halos identified in gravity-only cosmological simulations, along with statistical prescriptions to paint galaxies into these halos. 

Algorithms for assigning single band optical luminosities (or stellar masses) to mock galaxies using HOD, CLF or SHAM prescriptions have existed for about two decades \citep{cs02,vo04,rwtb13}, with relatively recent extensions to include colours (or star formation rates) \citep{ss09,hw13,caz21}. While the simplest occupation models single out halo mass as the primary driver of observed correlations between galaxy properties and their environments \citep[e.g.,][]{as07,zm15,phs18b,azpm19}, recent work has argued for the importance of modelling beyond-mass effects such as assembly bias and galactic conformity \citep{zhv14}, leading to tune-able prescriptions for these effects \citep{masaki+13,pkhp15,hearin+16,yuan+18,xzc21,caz21}. State-of-the-art implementations such as the \textsc{UniverseMachine} prescription of \citet{bwhc19} employ SHAM on entire merger trees in high-resolution gravity-only simulations, calibrated to reproduce stellar mass functions and star formation rates over a wide range of redshifts. 

On the \Hi\ side, mock catalogs have been created using a combination of galaxy formation SAMs and a prescription for distributing the \Hi\ in disks \citep[see, e.g.,][]{Obreschkow++2009}, which have been useful for planning upcoming surveys with telescopes such as the Square Kilometre Array (SKA).
If one is only interested in the very large-scale correlations of the \Hi\ distribution (e.g., for intensity mapping experiments), or in interpreting \Hi\ detections based on stacking experiments at high redshift, the algorithms for creating the catalogs are considerably simpler, assigning an \Hi\ mass directly to dark  halos using a physically motivated prescription \citep{Bagla+2010,GuhaSarkar++2012,Villaescusa-Navarro++2014,Castorina+2017,pra17}. 

In this  paper, we aim to combine low-redshift ($z\lesssim0.1$) constraints on galaxy abundances and clustering, from surveys of both optically selected as well as \Hi-selected galaxies, to generate mock galaxies that are simultaneously assigned multi-band optical information as well as \Hi\ masses. To this end, we consolidate recent work in these areas and introduce mock galaxy catalogs constructed using updated halo occupation models and calibration of multi-wavelength low-redshift galaxy optical and \Hi\ properties. Each mock galaxy in our catalogs is assigned values of the $r$-band absolute magnitude $M_r$ (detailed definition below), colour indices $g-r$ and $u-r$, \Hi\ mass \mHi\ and stellar mass $m_\ast$, along with a range of environmental properties derived from the dark matter environment of the galaxy's host halo. Relying on these properties, the mocks reproduce the luminosity function and the luminosity and colour dependence of projected 2-point clustering of SDSS galaxies, the \Hi\ mass function and \mHi-dependent clustering of massive ALFALFA galaxies, and predict the cross-correlations between these galaxies. The mocks also have tunable implementations of galactic conformity \citep{weinmann+06} and colour-dependent galaxy assembly bias \citep{hw13,pkhp15}.

Additionally, the host halo of each central galaxy in our mocks is `baryonified', i.e., assigned a realistic spatial distribution of stars and gas and, consequently, a realistic rotation curve for the galactic disk. This is a novel feature of our mocks which, as we discuss later, potentially allows us to explore a number of interesting questions that have not been adequately addressed in the theoretical literature. Among others, these include modelling the observed 21cm line profiles of \Hi-selected galaxies, along with the associated velocity width distribution, and the nature and universality of the radial acceleration relation in the baryons+cold dark matter (CDM) paradigm.

The rest of the paper is organised as follows. We describe the ingredients of our mock algorithm in section~\ref{sec:ingredients}, followed by a detailed description of the algorithm itself in section~\ref{sec:mocks}. In section~\ref{sec:results}, we demonstrate the performance of our mocks in reproducing a number of 1-point and 2-point statistical observables. In section~\ref{sec:predict_extend}, we discuss observables whose behaviour is \emph{predicted} by our mocks, along with possible extensions of our technique that are interesting for future analyses.  We conclude in section~\ref{sec:conclude} with a brief discussion of potential applications of our mocks.

Throughout, we consider a flat $\Lambda$CDM cosmology with parameters $\{\Omega_{\rm m},\Omega_{\rm b},h,n_{\rm s},\sigma_8\}$ given by $\{$0.276, 0.045, 0.7, 0.961, 0.811$\}$ compatible with the 7-year results of the \emph{Wilkinson Microwave Anisotropy Probe} experiment \citep[WMAP7,][]{Komatsu2010}, with a linear theory transfer function generated by the code \textsc{camb} \citep{camb}.\footnote{\url{http://camb.info}} Our convention will be to quote halo masses ($m$) in \Mh\ and galaxy stellar masses ($m_\ast$) and \Hi\ masses (\mHi) in \Mhsq\ units. The notation $m$ for halo mass will refer to $m_{\rm 200b}$, the mass enclosed in the radius $R_{\rm 200b}$ where the enclosed density falls to 200 times the background density. Similarly $m_{\rm vir}$ will refer to $m_{\rm 200c}$, the mass enclosed in the radius $R_{\rm 200c}$ where the enclosed density falls to 200 times the critical density.

\section{Ingredients}
\label{sec:ingredients}
We start by describing the ingredients used in constructing our mocks. These include the gravity-only simulations that we populate with galaxies, the observed galaxy sample whose optical properties form the basis of the halo occupation distribution (HOD) we use to assign luminosities, colours and stellar masses to the mock galaxies, and the scaling relation between optical properties and neutral hydrogen (\Hi) using which we  assign \Hi\ masses.

\subsection{Simulations}
\label{subsec:sims}
\noindent
The $N$-body simulations we rely on are listed in Table 1 of \citet{pa20}, of which we focus on the WMAP7 configurations. Specifically, we have 2, 10 and 3 realisations each of the ${\rm L}150\_{\rm N}1024$, ${\rm L}300\_{\rm N}1024$ and ${\rm L}600\_{\rm N}1024$ boxes, respectively, corresponding to particle masses $m_{\rm p}=2.41\times10^8,1.93\times10^9,1.54\times10^{10}\Mh$, respectively.
The notation ${\rm L}150\_{\rm N}1024$, for example, indicates a cubic, periodic box of length $L_{\rm box}=150\Mpch$ containing  $1024^3$ particles.
The simulations were performed using the tree-PM code \textsc{gadget-2} \citep{springel:2005}\footnote{\url{http://www.mpa-garching.mpg.de/gadget/}} with a PM grid of a factor $2$ finer than the initial particle count along each axis, and a comoving force softening length of $1/30$ of the mean interparticle spacing. 
Initial conditions were generated using $2^{\rm nd}$ order Lagrangian perturbation theory \citep{scoccimarro98} with the code \textsc{music}  \citep{hahn11-music}.\footnote{\url{https://www-n.oca.eu/ohahn/MUSIC/}} 
Halos were identified using the code \textsc{rockstar} \citep{behroozi13-rockstar}\footnote{\url{https://bitbucket.org/gfcstanford/rockstar}} which performs a Friends-of-Friends (FoF) algorithm in 6-dimensional phase space. We discard all sub-halos and further only consider objects whose `virial' energy ratio $\eta=2T/|U|$ satisfies $0.5\leq\eta\leq1.5$ \citep{Bett+07}. 
All the simulations and analysis were performed on the Perseus and Pegasus clusters at IUCAA.\footnote{\url{http://hpc.iucaa.in}}

\subsection{Galaxy sample}
\label{subsec:data}
\noindent
We rely on optical properties of galaxies in the local Universe as provided by Data Release 7 \citep[DR7,][]{abazajian+09} of the Sloan Digital Sky Survey \citep[SDSS,][]{york+00}.\footnote{\url{www.sdss.org}}
From the SDSS DR7 Catalog Archive Server (CAS),\footnote{\url{www.skyserver.sdss.org}} we obtained galaxy properties including Galactic extinction-corrected apparent magnitudes (luptitudes) in the $u$, $g$ and $r$ bands for all galaxies with spectroscopic redshifts in the range $0.02\leq z\leq 0.2$ and satisfying the Petrosian $r$-band apparent magnitude threshold $m_r\leq17.7$. Both Petrosian and Model magnitudes were obtained from the database. Absolute magnitudes $M_{{}^{0.1}u},M_{{}^{0.1}g},M_{{}^{0.1}r}$ were estimated by K-correcting to rest frame bands at $z=0.1$ using \textsc{K-correct} \citep{br07} (we used a modified version of the Python wrapper provided by N. Raseliarison\footnote{\url{https://github.com/nirinA/kcorrect\_python}}) and evolution correcting as described by \citet{blanton+03}. We did not correct for dust extinction in the host, which makes edge-on spirals appear redder. 
This makes our colour-dependent analysis consistent with that of \citet{zehavi+11} who reported measurements of colour-dependent clustering using similarly uncorrected colours. 
Correcting for inclination can, in principle, affect inferences regarding the physics of quenching in satellites, as well as the physics governing the \Hi\ content of optically red galaxies, which we will explore in future work.
(Consistency with \citealt{zehavi+11} is also why we do not work with the improved SDSS photometry discussed in \citealt{pymorph}.)
Flux measurement errors were accounted for when using \textsc{K-correct}, but not explicitly in the Gaussian mixture fitting below. 

This analysis yielded values of  $M_r\equiv M_{{}^{0.1}r}-5\log_{10}(h)$, and similarly $M_g$ and $M_u$, for each galaxy. The latter were converted to the colour indices $g-r = M_g-M_r$ and $u-r=M_u-M_r$, which are therefore rest frame colours, K-corrected and evolution corrected to $z=0.1$. Below, we use Petrosian absolute magnitudes $M_r$ and Model colours $g-r$ and $u-r$.

\subsection{Optical halo occupation distribution}
\label{subsec:opticalHOD}
\noindent
Here we discuss the complete optical HOD we use for assigning luminosities and colours to mock galaxies. This HOD is constrained by measurements of luminosity- and colour-dependent clustering in SDSS DR7 and uses a Gaussian mixture description of bi-variate colour distributions, as described below.

\subsubsection{Constraints from luminosity-dependent clustering}
\label{subsubsec:HODLum}
\noindent
We use the standard 5-parameter mass-only HOD calibrated by \citet{ppp19} for the WMAP7 cosmology, which describes SDSS luminosity-dependent clustering measurements from \citet{zehavi+11}. The HOD was calibrated using simulation-based tables of 2-point halo correlation functions and halo profiles from the $z=0$ outputs of the simulations described above, following the technique of \citet{zg16}. Satellites were assumed to be distributed according to the spherically averaged dark matter distribution in parent halos, without assuming the \citep[][NFW]{nfw96} profile (although the latter is an excellent approximation over the spatial and mass dynamic range of interest; this will be useful below). For the 2-halo terms, hard-sphere halo exclusion was implemented, but all other correlations were directly measured from the simulations in narrow mass bins, so that non-linear, scale-dependent halo bias was accurately modelled. See \citet{ppp19} for further discussion. 

\subsubsection{Constraints from colour dependent-clustering}
\label{subsubsec:HODCol}
\noindent
\citet{ppp19} also modelled colour-dependent clustering from \citet{zehavi+11} by treating the `red fraction' $f_{\rm r|s}(M_r)$ of satellites of luminosity $M_r$ as a free parameter. Here $f_{\rm r|s}(M_r)$ is the fraction of satellites in a luminosity bin (labelled $M_r$) whose $g-r$ colours satisfy 
\be
g-r > (g-r)_{\rm cut}(M_r) \equiv 0.21 - 0.03M_r\,,
\label{eq:gr-cut}
\ee
\citep[equation 13 of][]{zehavi+11}. \citet{ppp19} propagated this cut into the halo model formalism to observationally constrain $f_{\rm r|s}(M_r)$ in the four wide luminosity bins for which clustering measurements and jack-knife covariance matrices were available (see their Table 1). 

The final product provided by \citet{ppp19} comprises simple fitting functions for the $M_r$-dependence of the five parameters defining the thresholded HOD and the parameter $f_{\rm r|s}$ in the range $-23<M_r\leq-19$ (see their Figures 12 and 13 and Table 3; they denote $f_{\rm r|s}$ as $p_{\rm r|s}$). 
We use these below to assign luminosities and colours to mock central and satellite galaxies.

\begin{figure}
\centering
\includegraphics[width=0.45\textwidth]{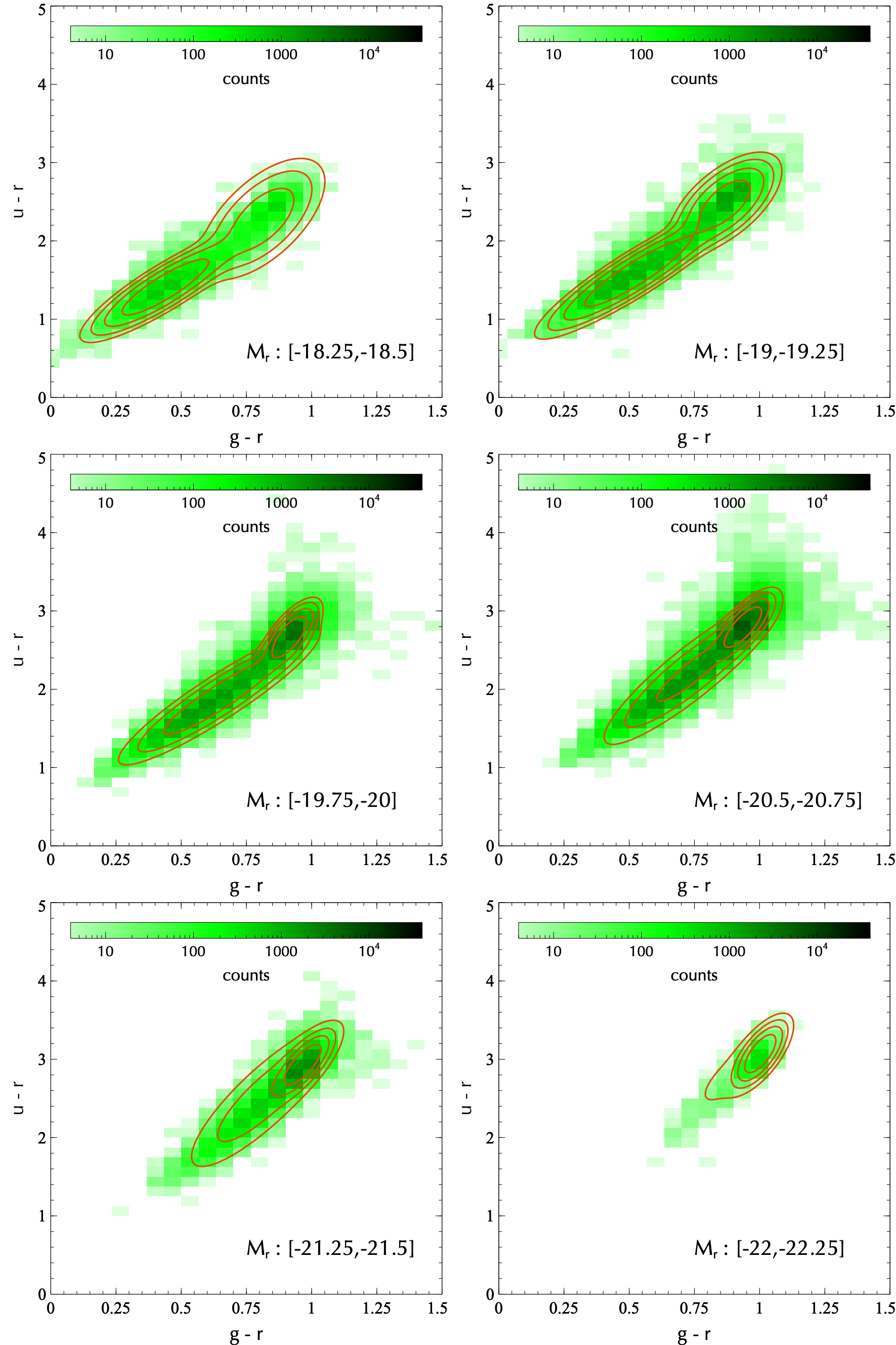}
\caption{Bivariate distribution of  Model $(g-r,u-r)$ colours in bins of Petrosian absolute magnitude $M_r$ (coloured histograms)  measured in SDSS, compared with the best fit Gaussian mixture in each bin (contours).}
\label{fig:gaussmix-comparedata}
\end{figure}

\begin{figure*}
\centering
\includegraphics[width=0.85\textwidth,trim=10 0 5 5,clip]{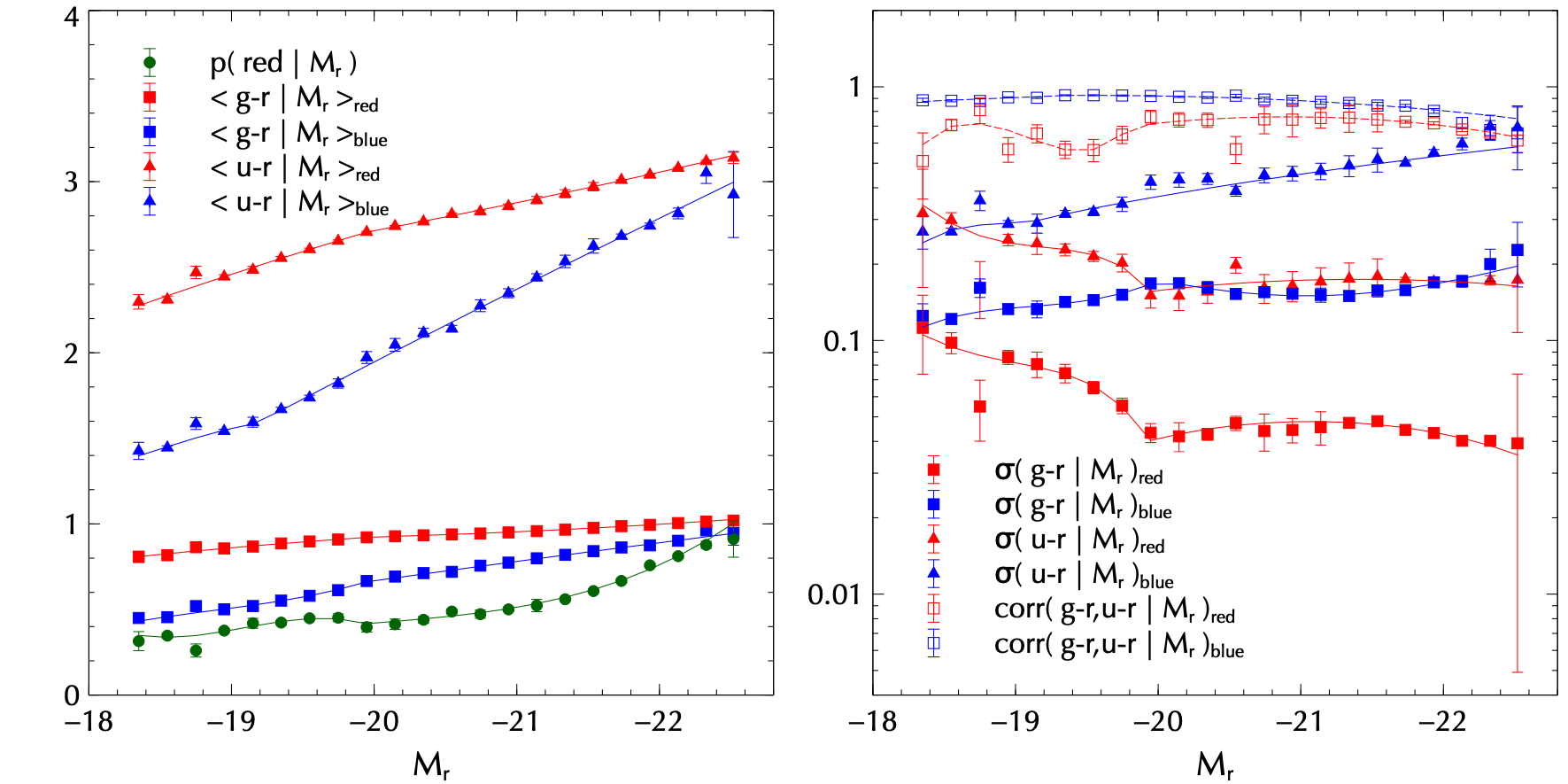}
\caption{Summary of Gaussian mixture fits for $p(g-r,u-r|M_r)$ (points with errors). Smooth curves show piece-wise continuous polynomial fits to each mixture parameter, which we use in assigning colours to mock galaxies.
}
\label{fig:gaussmix-fits}
\end{figure*}

\subsubsection{Colour distribution}
\label{subsubsec:dblGauss}
\noindent
In the following, we require the joint distribution $p(u-r,g-r|M_r)$ of $g-r$ and $u-r$ at fixed $M_r$ in SDSS, which we model as a 2-component bivariate Gaussian mixture. In principle, one can logically extend this analysis to multiple bands using a higher-dimensional Gaussian mixture. Likewise, we could use more components in the mixture, but have found that 2 are sufficient for the present purpose.  We use measured values of $g-r$ and $u-r$ in narrow bins of $M_r$ without accounting for measurement errors. For each bin of $M_r$, we construct a volume-limited subsample by choosing galaxies in the redshift range $0.02\leq z<z_{\rm max}(M_{r,{\rm max}})$, where $M_{r,{\rm max}}$ is the faint edge of the bin and $z_{\rm max}(M_{r,{\rm max}})$ is the redshift at which a galaxy of this absolute magnitude would fall below the flux limit $m_r=17.7$ of the survey. We use the Python package \texttt{sklearn.mixture} \citep{scikit-learn}\footnote{\url{https://scikit-learn.org/}} to implement an iterative Expectation-Maximisation algorithm \citep{dlr77} with 12 initialisations. We repeat the exercise in each luminosity bin for 150 bootstrap resamplings of the respective subsample and use the average values of the 11 parameters defining the Gaussian mixture as the `best fit', with the corresponding standard deviations across the bootstrap samples as errors.

Figure~\ref{fig:gaussmix-comparedata} compares the measured bivariate distributions in a few bins of $M_r$ (coloured histograms) with the best fit Gaussian mixture (contours). Figure~\ref{fig:gaussmix-fits} summarises the 11 parameters of the mixture for all the luminosity bins (points with errors). The parameters are: (i) the probability $p({\rm red}|M_r)$ that the galaxy belongs to the `red' mode of the mixture, (ii) the mean vectors $(\avg{g-r|M_r}_{\rm red},\avg{u-r|M_r}_{\rm red})$ and $(\avg{g-r|M_r}_{\rm blue},\avg{u-r|M_r}_{\rm blue})$ of the red and blue modes, respectively, (iii) the diagonal elements of the covariance matrices $(\sigma^2_{\rm red}(g-r|M_r),\sigma^2_{\rm red}(u-r|M_r))$ and $(\sigma^2_{\rm blue}(g-r|M_r),\sigma^2_{\rm blue}(u-r|M_r))$ of the red and blue modes, respectively and (iv) the correlation coefficients between $g-r$ and $u-r$, for the red and blue modes respectively. The smooth curves in Figure~\ref{fig:gaussmix-fits} show piece-wise continuous polynomial fits to each parameter. These are used similarly to the HOD fitting functions provided by \citet{ppp19}, to assign colours to mock galaxies (see below).

For assigning colours to centrals and satellites separately, we require the probability $p({\rm red}|{\rm sat},M_r)$ that the $g-r$ and $u-r$ colours of a satellite of luminosity $M_r$ are drawn from the `red mode' of the bivariate double-Gaussian distribution described above. This can be easily obtained by combining the constraints on $f_{\rm r|s}$ described in section~\ref{subsubsec:HODCol} with (a subset of) the parameters of the Gaussian mixture $p(g-r,u-r|M_r)$, and is given by
\begin{align}
p({\rm red}|{\rm sat},M_r) &= \frac{2f_{\rm r|s} - \Cal{I}_{\rm (blue)}(M_r)}{\Cal{I}_{\rm (red)}(M_r) - \Cal{I}_{\rm (blue)}(M_r)}
\label{eq:prs-from-frs}
\end{align}
where (suppressing the $M_r$-dependence of the arguments for brevity)
\be
\Cal{I}_{\rm (red/blue)}(M_r) = \erfc{\frac{(g-r)_{\rm cut}-\avg{g-r}_{\rm red/blue} }{ \sqrt{2}\sigma_{\rm red/blue}(g-r)}}\,,
\label{eq:Idef}
\ee
and where $(g-r)_{\rm cut}(M_r)$ was given in \eqn{eq:gr-cut}.

\subsubsection{Stellar masses}
\label{subsubsec:mstar}
\begin{figure}
\centering
\includegraphics[width=0.49\textwidth,trim=5 10 5 5,clip]{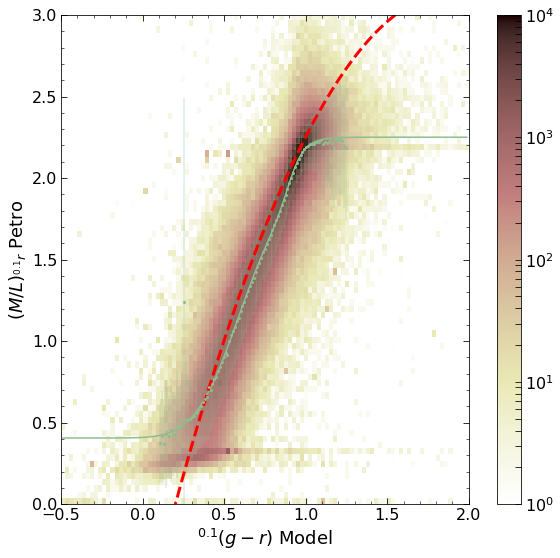}
\caption{Calibration of mass-to-light ratio for the SDSS DR7 sample, using the code \textsc{K-correct}. Histogram shows the measured distribution of Model $g-r$ colours against Petrosian mass-to-light ratio in the $r$-band.  Green points with errors show binned measurements of the same, and the green solid curve shows our best fit relation \eqref{eq:MbyL-mean} to the points. Dashed red curve shows the fit reported by \citet{ww12} for reference. See text for discussion. 
}
\label{fig:masstolight}
\end{figure}

\noindent
Stellar masses are calculated using a mass-to-light ratio calibrated to SDSS DR7 measurements similarly to \citet[][see their sections 2.3 and 4.2]{pkhp15}. The measurements are shown as the coloured histogram in Figure~\ref{fig:masstolight}. We fit a mean relation to these measurements in bins of $x\equiv (g-r)$, of the form
\be
\avg{M/L}(x) = a + b\,\erf{(x-c)d} + e\,\tanh((x-f)/g)\,,
\label{eq:MbyL-mean}
\ee
finding best fitting values
\begin{align}
a = 1.3281\,&;\quad b = 0.735\,; \quad c = 0.5859\,, \notag\\
d = 3.38\,&;\quad e = 0.187\,;\quad f = 0.8976\,;\quad g = 0.0874\,.
\label{eq:MbyL-mean-bestfit}
\end{align}
We also fit the scatter around the mean in the same bins, of the form
\be
\sigma_{(M/L)}(x) = \left\{ 
\begin{array}{ll}
  k_0 + k_1\,(x-x_0)   &;\, x < x_0  \\
  k_0 + k_2\,(x-x_0) + k_3\,(x-x_0)^2   & ;\, x \geq x_0\,,
\end{array}
\right.
\label{eq:MbyL-scat}
\ee
finding best-fitting values given by
\begin{align}
x_0 = 1.019\,&;\quad k_0 = 0.1441\,; \quad k_1 = -0.182\,, \notag\\
k_2 = 1.16\,&;\quad k_3 = -0.2\,.
\label{eq:MbyL-scat-bestfit}
\end{align}
Our calibration is shown in Figure~\ref{fig:masstolight} and is better behaved than that of \citet{pkhp15} for very blue objects; we have checked, however, that both calibrations lead to nearly identical results for the final mocks. For each mock galaxy with colour $g-r$, we calculate a Gaussian random number for the mass-to-light ratio with mean and standard deviation calculated using \eqns{eq:MbyL-mean} and \eqref{eq:MbyL-scat}, respectively, setting $x=(g-r)$.
This is then combined with the corresponding value of $M_r$ of the galaxy to assign a stellar mass $m_\ast$.

\subsection{Neutral hydrogen masses}
\label{subsec:mHI}
\noindent
\citet[][hereafter, PCP18]{pcp18} calibrated a lognormal scaling relation (with constant scatter in log-mass) between the neutral hydrogen mass \mHi\ of a galaxy and its optical properties $M_r$ and $g-r$, using the optical HOD from \citet{guo+15} and clustering measurements from \citet{guo+17} of \Hi-selected galaxies in the ALFALFA survey \citep{giovanelli+05}. The scaling relation was separately calibrated for central and satellite galaxies using a halo model, along with an overall parameter $f_{\Hi}$ which gives the fraction of optically selected galaxies that contain \Hi. Their default model also assumed that satellites with $\mHi > 10^{10.2}\times(0.678/h)^2\Msun$ do not exist. The model was constrained using measurements of number counts and projected clustering $w_{\rm p}(r_{\rm p})$ of ALFALFA \Hi-selected galaxies from \citet{guo+17} for the thresholds $\log_{10}(\mHi/\Mhsq) \geq 9.8+2\log_{10}(0.678)$ and $10.2+2\log_{10}(0.678)$.

The default PCP18 model  precludes rare, massive \Hi\ satellites.
In the following, we will  also see that this default implementation is in mild disagreement with $w_{\rm p}(r_{\rm p})$ for galaxies with $\log_{10}(\mHi/\Mhsq) \geq 10.0+2\log_{10}(0.678)$ (which was not used in constraining the scaling relation). This is likely due to our updated optical HOD from \citet{ppp19} with its improved modelling of scale-dependent halo bias and the self-consistent calibration of the satellite red fraction. (The latter was identified by PCP18 as the parameter most susceptible to systematic effects in their analysis; see their section~4.2.) We therefore explored variations around the PCP18 model, finding that the simple modification of excluding satellites from halos with $m > m_{\rm sat,max}$, while keeping the scaling relations for centrals and satellites intact otherwise, leads to acceptable descriptions of the measurements for all the thresholds mentioned above, as well as of the \Hi\ mass function. Since the threshold is on halo rather than \Hi\ mass, such a model can in principle accommodate rare and massive \Hi\ satellites. A simple $\chi^2$ minimisation exercise, varying $m_{\rm sat,max}$ and predicting $w_{\rm p}(r_{\rm p})$ for galaxies with $\log_{10}(\mHi/\Mhsq) \geq 10.0+2\log_{10}(0.678)$ using the covariance matrix kindly provided by Hong Guo, leads to $m_{\rm sat,max}\simeq10^{14.4}\Mh$.

Being tied to the optical HOD, this \Hi\ scaling relation suffers from a natural incompleteness in producing \Hi\ masses, determined by the optical completeness limit on $M_r$ for the galaxy population. For a sample limited by $M_r<-19$, for example, the \Hi\ mass function is complete only above $\mHi \gtrsim 10^{9.65}\Mhsq$. As such, all our analysis of \Hi-selected galaxies is restricted to the massive end. 

\section{Assigning galaxy properties in mocks}
\label{sec:mocks}
\noindent
We now describe our main algorithm for generating mock galaxies in a gravity-only simulation, along with a `baryonification' scheme for assigning spatial distributions of stars and gas to each galaxy.

\subsection{Algorithm}
\label{subsec:algorithm}
\noindent
Our basic algorithm to assign galaxy luminosities and colours to halos in $N$-body simulations is essentially the same as that described by \citet{ss09}, with a few technical improvements. We also include a number of additional galaxy properties. The algorithm can be summarised as follows:
\begin{itemize}
\item \underline{\bf Central occupation and luminosity:} A threshold luminosity $L_{\rm min}$ (or absolute magnitude $M_{r,{\rm max}}$) is dynamically determined by requiring that the HOD of central galaxies $f_{\rm cen}(>L_{\rm min}|m)$ be sampled down to a value $1.5\times10^{-3}$ for all halo masses $m\geq40\,m_{\rm part}$, where $m_{\rm part}$ is the particle mass of the simulation. For example, for the ${\rm L}300\_{\rm N}1024$ configuration, this gives us $M_{r,{\rm max}}\simeq-19$. halos of mass $m$ are then occupied by a central galaxy with probability $f_{\rm cen}(>L_{\rm min}|m)$. Central galaxy luminosities are sampled from the conditional luminosity function $f_{\rm cen}(>L|m)/f_{\rm cen}(>L_{\rm min}|m)$.
\item \underline{\bf Satellite occupation and luminosity:} The number of satellites in an occupied halo is drawn from a Poisson distribution with mean $\bar N_{\rm sat}(>L_{\rm min}|m)$, and the luminosities of these satellites are sampled from the conditional luminosity function $\bar N_{\rm sat}(>L|m)/\bar N_{\rm sat}(>L_{\rm min}|m)$. 
We do not enforce that satellites be less luminous than their host central. 
For a luminosity-complete sample of galaxies with $M_r\leq-19$, we find that, in approximately $6.5\%$ of groups containing at least one satellite, the brightest satellite is brighter than the central.
\item \underline{\bf Galaxy positions:}  The central of a halo is placed at the halo center-of-mass. Satellite positions are distributed as an NFW profile around the central, truncated at $R_{\rm 200b}$. Halo concentrations $c_{\rm 200b}$ are drawn from a Lognormal distribution with median and scatter at fixed halo mass as calibrated by \citet{dk15} using very high-resolution simulations, which avoids contamination due to numerical fitting errors in relatively low-resolution boxes. 
The use of the NFW form, which is an accurate description of the profile of well-resolved halos at the masses and spatial separations of our interest \citep[see, e.g., Figures~2 and~3 of][]{ppp19}, also allows us to bypass the  need for accessing particle information from the underlying $N$-body simulation and work only with halo catalogs.
The fitting functions from \citet[][who used the $m_{\rm 200c}$ definition of halo mass]{dk15} are converted to the $m_{\rm 200b}$ definition appropriate for the HOD fits using the prescription of \citet{hk03}. 

To preserve correlations between halo concentration and large-scale environment, the Lognormal concentrations of halos in narrow mass bins are rank ordered by the actual estimated halo concentrations in each bin \citep[the assumption being that this ranking would be approximately preserved even in relatively coarsely sampled halos, although see][]{rps21}. The scale radius $r_{\rm s} = R_{\rm 200b}/c_{\rm 200b}$ inferred for each halo from this exercise is stored for later use (see section~\ref{subsec:baryonify}).
\item \underline{\bf Galaxy velocities:} The central of a halo is assigned the bulk velocity of the halo. The satellites are assigned random velocities drawn from a 3-dimensional isotropic Gaussian distribution with mean equal to the central velocity and 1-d velocity dispersion appropriate for the NFW profile at the location of the satellite.
\end{itemize}

Additionally, we implement the following modifications:
\begin{itemize}
\item \underline{\bf Galaxy colours:} To assign $g-r$ and $u-r$ colours to galaxies, we extend the algorithm proposed by \citet{ss09}. As in their case, the first step is to determine whether a galaxy is `red' or `blue', which is done separately for satellites and centrals, using the probability $p({\rm red}|{\rm sat},M_r)$ for satellites (see sections~\ref{subsubsec:dblGauss}) and the corresponding probability for centrals which can be derived using $p({\rm red}|{\rm sat},M_r)$, $p({\rm red}|M_r)$ and the HOD \citep[this also requires an integral over the halo mass function, for which we use the fitting function from][]{Tinker08}. In the next step, we assign $g-r$ and $u-r$ colours by sampling the appropriate (bivariate) mode of the double Gaussian $p(g-r,u-r|M_r)$ calibrated in section~\ref{subsubsec:dblGauss}. Operationally, we first sample the univariate mode $p_{\rm red/blue}(g-r|M_r)$ (obtained by marginalising the respective bivariate distribution over $u-r$) and then sample the corresponding conditional distribution $p_{\rm red/blue}(u-r|g-r,M_r)$, for each red/blue galaxy \citep[see also][]{xu+18}. Optionally, we also include galactic conformity by correlating $g-r$ with halo concentration (or an unspecified Gaussian-distributed halo property) at fixed halo mass, using the tunable prescription of \citet{pkhp15}.
\item \underline{\bf Galaxy stellar masses:} As discussed by \citet{pkhp15} and described in detail in section~\ref{subsubsec:mstar}, we calculate stellar masses $m_\ast$ using a $(g-r)$-dependent mass-to-light ratio calibrated for SDSS DR7 galaxies.
\item \underline{\bf Galaxy neutral hydrogen masses:} We assign \Hi\ masses to a uniformly sampled fraction $f_{\Hi}$ of galaxies, with a lognormal distribution at fixed $M_r$ and $g-r$, using the model from PCP18. This model, which we refer to as the `minimal PCP18' model below, additionally discards \Hi-satellites which have $\log_{10}(\mHi/\Mhsq) > 10.2 + 2\log_{10}(0.678)$. As described in section~\ref{subsec:mHI}, we modify this model by instead discarding \Hi-satellites in parent halos having $m \geq m_{\rm sat,max}$.\footnote{All such `discarded' satellites continue to have their assigned optical properties, only their \Hi\ mass is set to zero.} This is done \emph{after} the uniform downsampling for \Hi\ assignment described above. For our default model, which we call `PCP18 mod-sat', we set $m_{\rm sat,max}=10^{14.4}\Mh$ (see below for a comparison between the models). 
As discussed by PCP18, for optically selected galaxies with $M_r\leq-18$, this optical-\Hi\ scaling relation places the majority of \Hi\ mass in faint blue galaxies (see their Figure~9).
\item \underline{\bf Neutral hydrogen disks:} For each galaxy containing \Hi, we assign a comoving disk scale length $h_{\Hi}$ for an assumed thin disk with \Hi\ surface density $\Sigma_{\Hi}(r_\perp)\propto\e{-r_\perp/h_{\Hi}}$ (here $r_\perp$ is the radial distance in the disk plane), using the scaling relation 
\be
h_{\Hi} = 7.49\,\kpch \left(m_{\Hi}/10^{10}\Mhsq\right)^{0.5}\,,
\label{eq:hdisk}
\ee
with a scatter of $0.06$ dex, consistent with the measurements reported by \citet{wang+16-HI}.\footnote{\citet{wang+16-HI} provide a scaling relation for the quantity $D_{\Hi}$ defined as the diameter of the contour corresponding to a surface density of $1\Msun\,{\rm pc}^{-2}$, which we relate to $h_{\Hi}$ using the provided scaling relation itself (which implies a constant average surface density, independent of $m_{\Hi}$) along with an integral over the exponential disk profile $\Sigma_{\Hi}(r_\perp)$.} These are useful in modelling `baryonified' rotation curves, as we describe later.
\item \underline{\bf Galaxy environment:} We assign to each galaxy the value of its host-centric dark matter overdensity $\delta$ and tidal anisotropy $\alpha$, each Gaussian smoothed at an adaptive scale $4R_{\rm 200b,host}/\sqrt{5}$ and at the fixed scale $2\Mpch$, as well as the bias $b_1$ of the host (see \citealp{phs18a} and \citealp{pa20} for how these variables are calculated in the $N$-body simulation). 
\end{itemize}

\subsection{`Baryonified' density profiles and rotation curves}
\label{subsec:baryonify}
\noindent
The galaxy properties described above, namely, luminosity, colour, stellar mass and \Hi\ mass, are all assigned by our algorithm by treating each galaxy as a point object. Combined with the information on the host halo mass and concentration, however, these properties can also be used to model the \emph{spatial distribution} of stars and of hot and cold gas in the galaxy and its halo. This in turn can be used to construct a rotation curve for the galaxy, which has several interesting applications as we discuss later. 

Since the circular velocity $v_{\rm rot}(r)$ at a halo-centric distance $r$ depends on the total mass $m_{\rm tot}(<r)$ enclosed in this radius, we must model the spatial distribution of \emph{all} matter components inside the host halo. This is particularly relevant for the inner parts of the halo which are typically baryon-dominated.
We restrict this analysis to central galaxies and will return in future work to satellite galaxies, which require additional modelling of processes such as tidal and ram pressure stripping, strangulation, etc. \citep[see, e.g.,][]{vdb+08,bwhc19} that are beyond the scope of the present work.

We follow the prescription of \citet[][henceforth, ST15]{st15} to `baryonify' each host halo. This method, and extensions thereof, have been shown to successfully account for baryonic effects in the matter power spectrum \citep{chisari+18,schneider+19,arico+20a} and bispectrum \citep{arico+20b} at relatively small scales over a range of redshifts. We have modifed the ST15 prescription to include the \Hi\ disk and have simplified it by truncating all profiles at the halo radius \citep[see, e.g.,][]{arico+20a}. Following the general practice for this method, we use the mass $m_{\rm vir}\equiv m_{\rm 200c}$ and radius $R_{\rm vir}\equiv R_{\rm 200c}$ for all baryonification scaling relations below. We are primarily interested here in low redshifts and length scales $\lesssim R_{\rm vir}$. We briefly summarise the method next. 

\begin{figure*}
\centering
\includegraphics[width=0.45\textwidth,trim=8 5 5 0,clip]{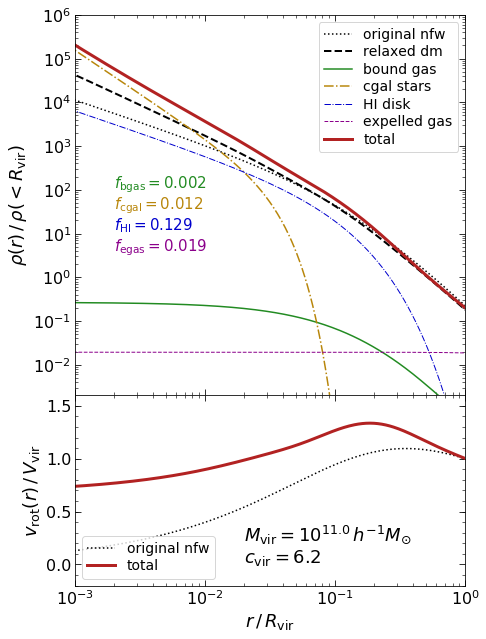}
\includegraphics[width=0.45\textwidth,trim=8 10 5 10,clip]{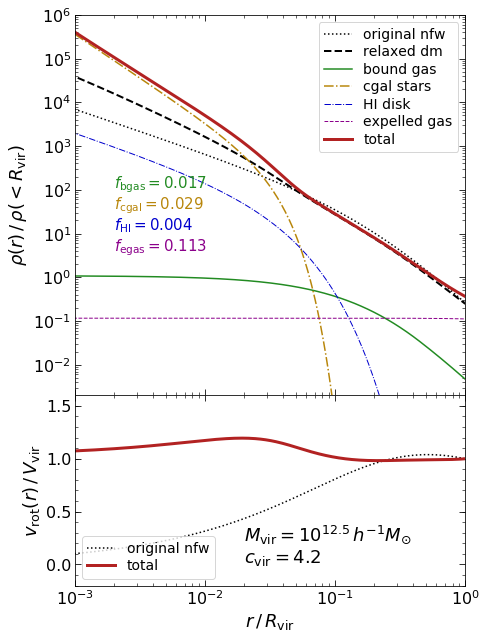}
\caption{Baryonification of halos from gravity-only simulations. We show two examples of hypothetical halos with masses $m_{\rm vir}=10^{11}\Mh$ \emph{(left panels)} and $m_{\rm vir}=10^{12.5}\Mh$ \emph{(right panels)}, with halo concentrations and baryonic mass fractions as indicated in respective labels. The baryonic fractions sum up to $\Omega_{\rm b}/\Omega_{\rm m} \simeq 0.163$. \emph{Upper panels:} Density profiles normalised by the halo density $\rho(<R_{\rm vir}) \equiv 200\rho_{\rm crit}$. The original NFW profile in each case is shown by the thin dotted black curve. Thicker curves with different colours and line styles show the 5 individual components as indicated, with the thick dark red curve showing the final total profile. Note especially that the ejected gas profile (dotted magenta) is nearly a constant in each case, and that the relaxed dark matter profile (thick dashed black) is substantially different from the original NFW in its shape due to quasi-adiabatic contraction and expansion. \emph{Lower panels:} Rotation velocity profile $v_{\rm rot}(r)$ normalised by the virial velocity $V_{\rm vir} = \sqrt{Gm_{\rm vir}/R_{\rm vir}}$. 
Note that the contribution of the \Hi\ disk to the rotation curve is treated separately from that of the spherical components, as described in the text. }
\label{fig:baryonification}
\end{figure*}

\begin{itemize}
\item Before baryonification, each halo starts with its total matter as a single component distributed according to an NFW profile consistent with the gravity-only simulation in which the mock catalog is being generated, using the mass $m_{\rm vir}$ and concentration $c_{\rm vir}=R_{\rm vir}/r_{\rm s}$ (see section~\ref{subsec:algorithm} for details of determining the scale radius $r_{\rm s}$ for each halo).
\item We divide the total mass of each baryonified halo into 5 components: bound gas (`bgas'), stars in the central galaxy  (`cgal'), neutral hydrogen in its disk (`\Hi'),\footnote{We assume that the stellar and \Hi\ disks are decoupled and do not model time-dependent warps, etc. in the \Hi\ disk. We correct for Helium as mentioned in the text but do not attempt to account for molecular Hydrogen.} gas expelled due to feedback (`egas') and the dark matter which quasi-adiabatically relaxes in the presence of the baryons (`rdm').
\item Each baryonic component, denoted by index $\alpha\in\{\rm bgas,\,cgal,\,\Hi,\,egas\}$, is assigned a mass fraction $f_\alpha$ subject to the constraint $f_{\rm bary}\equiv\sum_\alpha\,f_\alpha=\Omega_{\rm b}/\Omega_{\rm m}$ due to conservation of baryonic mass. In practice, we set $f_{\rm cgal}=m_\ast/m_{\rm vir}$, $f_{\Hi}=1.33\,m_{\Hi}/m_{\rm vir}$ (with the prefactor accounting for Helium correction) and $f_{\rm bgas}$ using
\be
f_{\rm bgas} = (\Omega_{\rm b}/\Omega_{\rm m})\times\left[1+(M_{\rm c}/m_{\rm vir})^\beta\right]^{-1}\,,
\label{eq:fbgas}
\ee
with $M_{\rm c}=1.2\times10^{14}\Mh$ and $\beta=0.6$ as described by ST15 (see their equation 2.19 and figure~2). We comment on possible variations in this relation later.
The ejected gas fraction $f_{\rm egas}$ is then set by the baryonic mass conservation constraint.\footnote{For a small fraction ($\sim1\%$) of objects with $M_r\leq-19$, the sum $f_{\rm cgal} + f_{\Hi} + f_{\rm bgas}$ exceeds $\Omega_{\rm b}/\Omega_{\rm m}$ (these in turn are dominated by objects having $f_{\rm cgal} + f_{\Hi} > \Omega_{\rm b}/\Omega_{\rm m}$). For such objects, we set $f_{\rm egas}=0$ without changing any of the other baryonic mass fractions, so that $f_{\rm bary} >\Omega_{\rm b}/\Omega_{\rm m}$. Overall mass conservation then implies that the corresponding dark matter fraction $f_{\rm rdm}=1-f_{\rm bary}$  is smaller than $1-\Omega_{\rm b}/\Omega_{\rm m}$ for these objects.}
\item Each baryonic component is given its own mass profile $\rho_\alpha(r)$. The choices below for $\rho_{\rm bgas}$, $\rho_{\rm cgal}$ and $\rho_{\rm egas}$ are identical to those in ST15. We briefly describe these below and refer the reader to ST15 for more details and original references.
\begin{itemize}
\item The bound gas component refers to the hot, ionized halo gas (which does not include, e.g., gas heated by supernovae). The corresponding density 
profile $\rho_{\rm bgas}$ has the form $\rho_{\rm bgas}\propto \left[\ln(1+r/r_{\rm s})/(r/r_{\rm s})\right]^{1/(\Gamma-1)}$,  set assuming hydrostatic equilibrium and a polytropic equation of state in the inner halo and matched to the original NFW profile in the outer halo. Here $r_{\rm s}$ and $\Gamma$ are the scale radius of the NFW profile and the polytropic index of the gas, respectively. The matching is performed at a radius $r_{\rm match}=\sqrt{5}\,r_{\rm s}$ and fixes the value of $\Gamma$. For hosts of centrals with $M_r\leq-19$, we find typical values of $\Gamma\simeq1.19$ with a dispersion of $\simeq0.015$.
\item The stellar profile is assumed to follow $\rho_{\rm cgal}\propto r^{-2}\,\e{-r^2/4R_{\rm hl}^2}$ with half-light radius $R_{\rm hl}=0.015\,R_{\rm vir}$ \citep{kravtsov13}. In principle, this can be extended to include a scatter and/or accommodate a dependence on halo angular momentum as predicted by disk formation models \citep[][see the discussion in \citealp{kravtsov13}]{mmw98}; we ignore this here for simplicity. Strictly speaking, we should treat the stellar profile as a combination of a central bulge and a 2-dimensional disk (with a relative contribution that correlates with galaxy colour), rather than the purely spherically symmetric form assumed here. 
In this work, we will follow the previous literature on the subject and assume the form given above, leaving a more self-consistent description of the stellar  disk to future work. In this sense, the stellar profile we model is better thought of as a pure bulge. 
\item The expelled gas component incorporates all gas affected by feedback, without distinguishing between possible gas phases. Its
profile is taken to be $\rho_{\rm egas}(r)\propto \e{-r^2/2r_{\rm ej}^2}$ with $r_{\rm ej}=0.5\sqrt{200}\,\eta_{\rm ej}\,R_{\rm vir}$, setting $\eta_{\rm ej}=0.5$, so that $r_{\rm ej}\simeq3.5R_{\rm vir}$. As discussed by ST15, the modelling of $\rho_{\rm egas}(r)$ in the halo outskirts is rather uncertain and observationally ill-constrained. However, at the scales of our interest ($r\lesssim R_{\rm vir}$), $\rho_{\rm egas}(r)\approx$ constant and its contribution to the total mass is therefore completely determined by baryonic mass conservation inside the halo. Our results are therefore expected to be very robust to any minor variations in the shape of $\rho_{\rm egas}(r)$ at scales $\gtrsim R_{\rm vir}$.
\item The sphericalised profile of \Hi\ is obtained by integrating a thin exponential disk of surface density $\Sigma_{\Hi}(r_\perp)\propto\e{-r_\perp/h_{\Hi}}$ (with $r_\perp$ being the radial distance in the disk plane) to get $\rho_{\Hi}(r)\propto\,r^{-1}\,\e{-r/h_{\Hi}}$. The disk scale length $h_{\Hi}$ is assigned as described in section~\ref{subsec:algorithm}. {\bf Note:} This sphericalised contribution only affects the calculation of the relaxed dark matter component. The contribution of the \Hi\ disk to  the rotation curve itself is treated separately as described below.
\end{itemize}

\begin{figure*}
\centering
\includegraphics[width=0.45\textwidth]{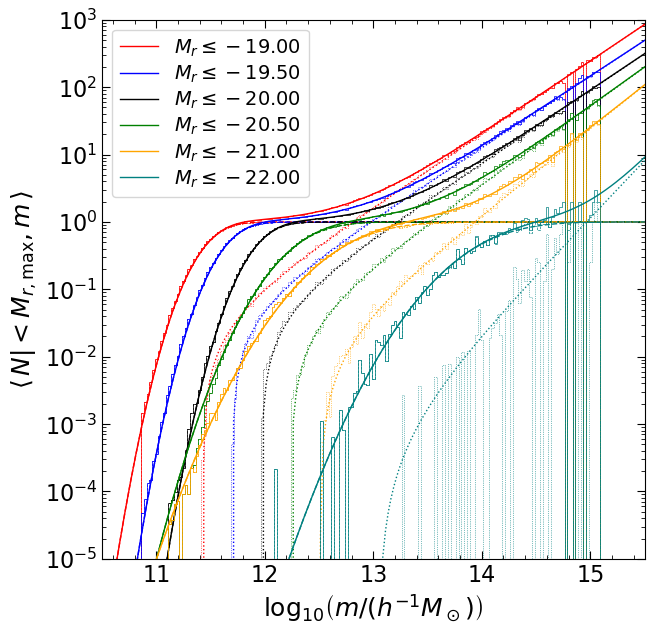}
\includegraphics[width=0.45\textwidth]{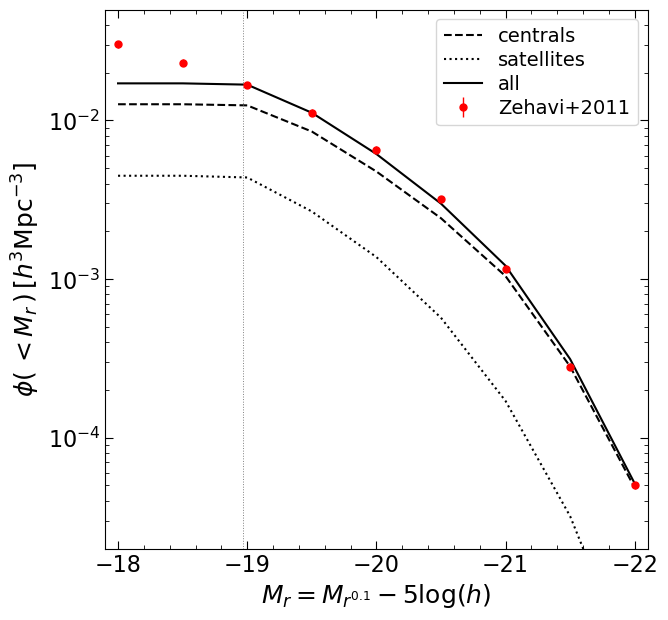}
\caption{Sanity check on HOD. \emph{(Left panel:)} Mocks versus input functions. For each threshold on $M_r$, we separately show the contribution of central and satellite galaxies in the mock (histograms) and in the analytical HOD (smooth curves). \emph{(Right panel:)} Thresholded luminosity function in the mock (separately showing the contribution of central, satellite and all galaxies) versus SDSS data from \citet{zehavi+11}. Vertical dotted line indicates the completeness threshold calculated by our algorithm for the ${\rm L}300\_{\rm N}1024$ box.}
\label{fig:hodlumfunc}
\end{figure*}

\begin{figure}
\centering
\includegraphics[width=0.23\textwidth]{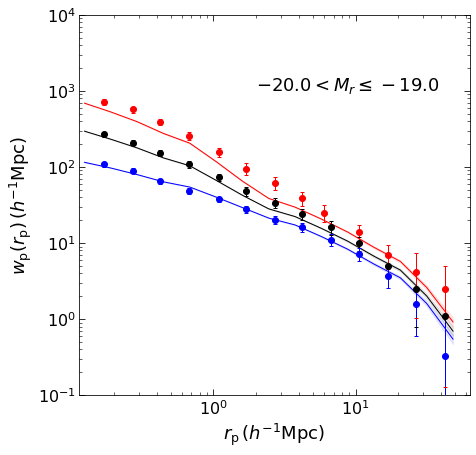}
\includegraphics[width=0.23\textwidth]{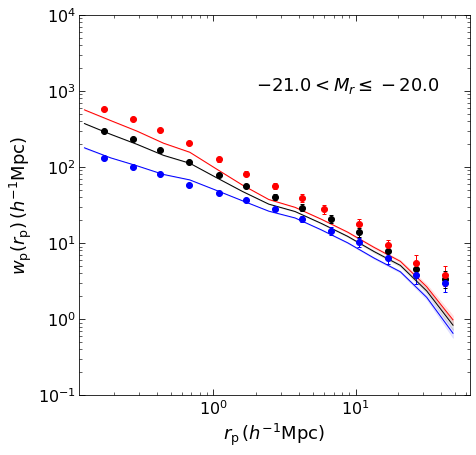}\\
\includegraphics[width=0.23\textwidth]{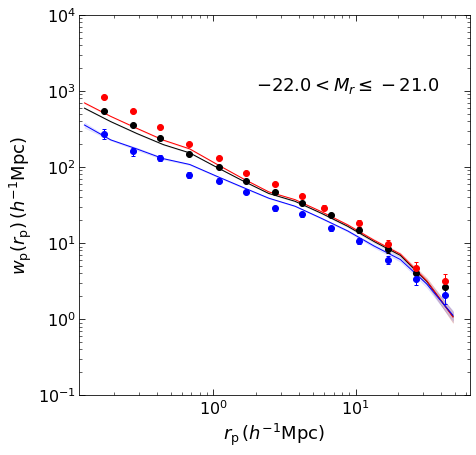}
\includegraphics[width=0.23\textwidth]{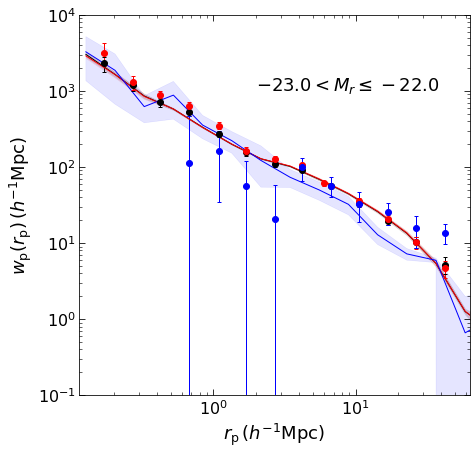}
\caption{Sanity check on clustering. Projected 2-point correlation function (2pcf) $w_{\rm p}(r_{\rm p})$ for red/blue/all galaxies in luminosity bins in the mock (solid lines with error bands) compared with data from \citet[][points with errors]{zehavi+11}. Mock measurements used 6 realisations of the ${\rm L}300\_{\rm N}1024$ box for the three fainter bins and 3 realisations of the ${\rm L}600\_{\rm N}1024$ box for the brightest bin. Lines show the mean and error bands reflect the respective standard deviations over all available realisations. Similarly to the data, mock galaxies were classified as red and blue based on their $g-r$ values in comparison to \eqn{eq:gr-cut}. The 2pcf for mock galaxies was calculated using \eqn{eq:wprp} with $\pi_{\rm max}=60\Mpch$ to match the \citet{zehavi+11} measurements.}
\label{fig:wprp}
\end{figure}

\noindent
Each profile function $\rho_\alpha(r)$ is normalised so as to enclose the \emph{total} mass $m_{\rm vir}$ inside $R_{\rm vir}$. The total baryonic profile is then $f_{\rm bary}\,\rho_{\rm bary}(r) = \sum_\alpha\,f_\alpha\,\rho_\alpha(r)$, with an enclosed baryonic mass $m_{\rm bary}(<r)=4\pi\int_0^r\der r^\prime\,r^{\prime2}\,f_{\rm bary}\,\rho_{\rm bary}(r^\prime)$.
\item The dark matter component is assumed to quasi-adiabatically respond to the presence of baryonic mass and relax to a new shape while approximately conserving angular momentum. The details of the iterative procedure used to calculate the resulting relaxed dark matter profile $\rho_{\rm rdm}(r)$ (normalised similarly to the baryonic profile functions) are in Appendix~\ref{app:rdm}, which is based on section 2.3 of ST15. 
The mass fraction $f_{\rm rdm}$ is set simply by mass conservation to be $f_{\rm rdm} = 1 - \Omega_{\rm b}/\Omega_{\rm m}$. The  dark matter mass enclosed in radius $r$ is then $m_{\rm rdm}(<r) = 4\pi\int_0^r\der r^\prime\,r^{\prime2}\,f_{\rm rdm}\,\rho_{\rm rdm}(r^\prime)$.
\item The thin exponential \Hi\ disk leads to a mid-plane circular velocity contribution $v_{\Hi}(r)$ satisfying  \citep[see section 2.6 of][]{binney-tremaine-GalDyn}
\be
v_{\Hi}^2(r) = \frac{2f_{\Hi}V_{\rm vir}^2}{\left(h_{\Hi}/R_{\rm vir}\right)}\,y^2\left[I_0(y)K_0(y) - I_1(y)K_1(y)\right]\,,
\label{eq:expdisk-vrot}
\ee
where $y\equiv r/(2h_{\Hi})$, $V_{\rm vir} = \sqrt{Gm_{\rm vir}/R_{\rm vir}}$ is the virial velocity and $I_n(y)$ and $K_n(y)$  are  modified Bessel functions of the first and second kind, respectively.
\item The total mass of dark matter and all baryonic components except the \Hi\ disk enclosed in radius $r$ is $m_{\rm tot-\Hi}(<r)=m_{\rm rdm}(<r)+m_{\rm bary-\Hi}(<r)$, where the notation `bary-\Hi' refers to summing over all baryonic components except \Hi. Since we truncate all profiles at the radius $R_{\rm vir}$ of the host halo, the total halo mass satisfies $m_{\rm vir}=m_{\rm tot-\Hi}(<R_{\rm vir})+m_{\Hi}$. Finally, the rotational velocity $v_{\rm rot}(r)$ of a test particle at halo-centric distance $r$ in the galaxy mid-plane is 
\be
v_{\rm rot}(r)=\sqrt{Gm_{\rm tot-\Hi}(<r)/r + v_{\Hi}^2(r)}\,.
\label{eq:vrot}
\ee
\end{itemize}
Figure~\ref{fig:baryonification} shows baryonified density profiles and rotation curves for hypothetical dwarf-like and Milky Way-like halos hosting an NGC 99-like galaxy, comparing the baryonified result with the original NFW result in each case. Only for these examples, we have set the galaxy stellar mass using the abundance matching prescription of \citet{behroozi+13-SHAM} with updated parameters taken from \citet{kvm18}, and have fixed the \Hi\ mass to $m_{\Hi} = 10^{9.83}\Mhsq$, with the \Hi\ disk size set using \eqn{eq:hdisk}.

We clearly see that the ejected gas profile is essentially constant in each case. More importantly, we see that the baryonified rotation curves are substantially flatter and also more diverse in shape than their purely NFW counterparts (see also section~\ref{subsec:galDMconnection}).

\begin{figure*}
\centering
\includegraphics[width=0.99\textwidth,trim = 10 8 10 5,clip]{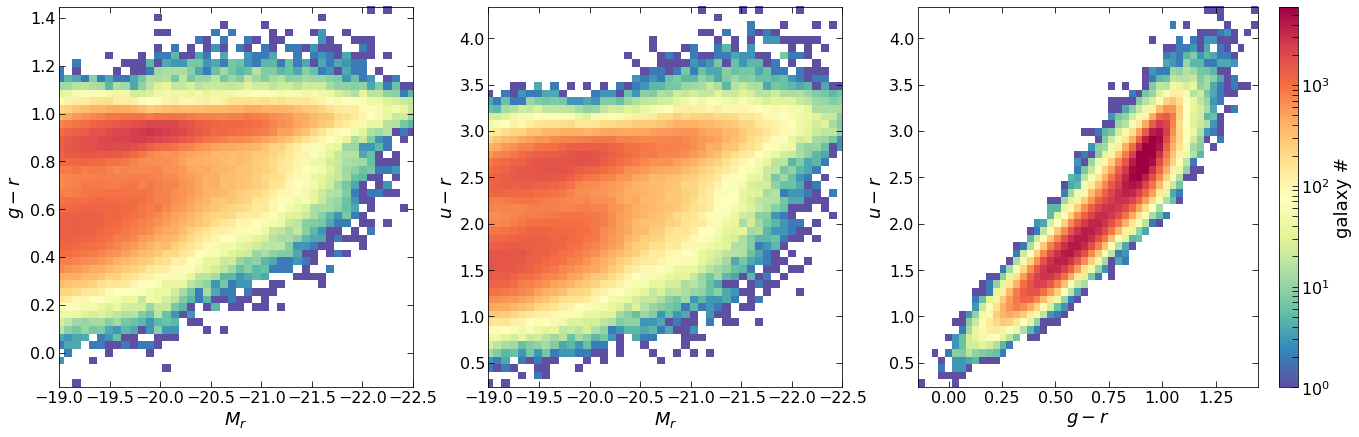}
\caption{Colour-magnitude and colour-colour bimodality in the mock. Histograms show distributions of $M_r$ against $g-r$ \emph{(left panel)} and $u-r$ \emph{(middle panel)}, and $g-r$ against $u-r$ \emph{(right panel)} in one mock using the ${\rm L}300\_{\rm N}1024$ box.}
\label{fig:colmag}
\end{figure*}

\section{Results}
\label{sec:results}
\noindent
We now report the results of generating mock catalogs using the algorithm of section~\ref{sec:mocks} on the simulations described in section~\ref{subsec:sims}.

\begin{figure*}
\centering
\includegraphics[width=0.45\textwidth]{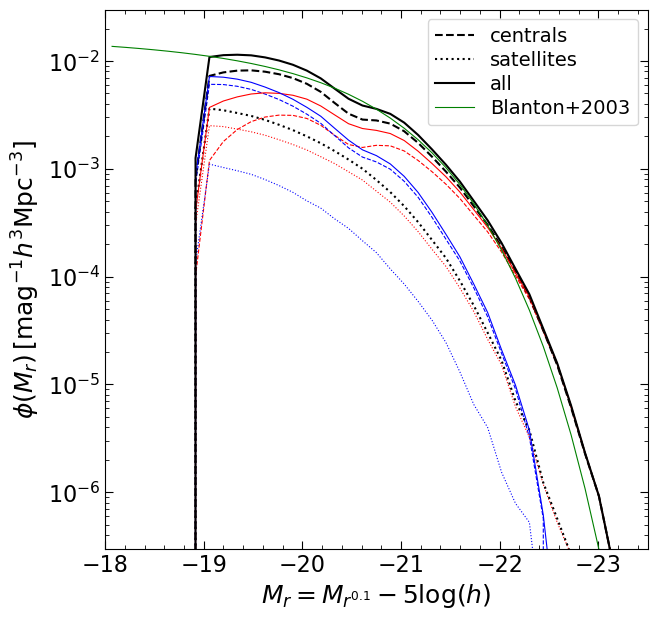}
\includegraphics[width=0.45\textwidth]{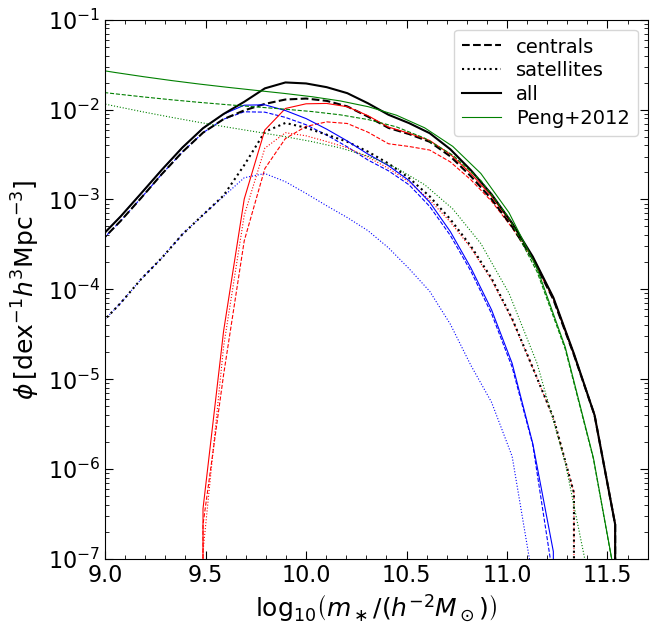}
\caption{Differential luminosity functions \emph{(left panel)} and stellar mass functions \emph{(right panel)} of red/blue/all central/satellite/all galaxies averaged over 6 mocks using the ${\rm L}300\_{\rm N}1024$ configuration. Colour segregation in the \emph{left panel} was based on \eqn{eq:gr-cut} applied to $g-r$ and $M_r$ for each galaxy, while for the \emph{right panel} we used $(g-r)_{\rm cut}=0.76 + 0.10[\log_{10}(m_\ast/\Mhsq)-10]$ \citep{pkhp15}. For comparison, the solid green curves in the \emph{left (right) panel} show the corresponding SDSS fits from \citet{blanton+03} \citep{peng+12}. In the case of the stellar mass function, these fits are also separately available for centrals and satellites and are shown as the dashed and dotted green curves, respectively}
\label{fig:Mrm*func}
\end{figure*}

\subsection{Optical properties}
\label{subsec:optical}
\noindent
As a sanity check, the \emph{left panel} of Figure~\ref{fig:hodlumfunc} compares the input fitting functions for the HOD from \citet{ppp19} with the output of the mock algorithm applied to a single ${\rm L}300\_{\rm N}1024$ box. The \emph{right panel} of the Figure compares the thresholded luminosity function averaged over 6 realisations of the ${\rm L}300\_{\rm N}1024$ box with the measurements from \citet{zehavi+11} that were used as constraints by \citet{ppp19}. 

Figure~\ref{fig:wprp} similarly compares the projected clustering of red/blue/all galaxies in mock catalogs with the measurements from \citet{zehavi+11}. Mock galaxies were classified as red and blue based on their $g-r$ values in comparison to \eqn{eq:gr-cut} to ensure a fair comparison with the data. The projected 2-point correlation function (2pcf) $w_{\rm p}(r_{\rm p})$ for mock galaxies was calculated by integrating the real space 2pcf $\xi(r)$ using
\be
w_{\rm p}(r_{\rm p}) = 2\int_{r_{\rm p}}^{\sqrt{r_{\rm p}^2+\pi_{\rm max}^2}}\der r\,\frac{r\,\xi(r)}{\sqrt{r^2-r_{\rm p}^2}}\,,
\label{eq:wprp}
\ee
where we set $\pi_{\rm max}=60\Mpch$ to match the \citet{zehavi+11} measurements.
We see generally good agreement in all cases, although the clustering of red galaxies tends to be lower in the mock than in the data. This is very likely due to the limited volume of our $300\Mpch$ boxes, which do not include the effects of faint (predominantly red) satellites in very massive halos.

\begin{figure*}
\centering
\includegraphics[width=0.45\textwidth]{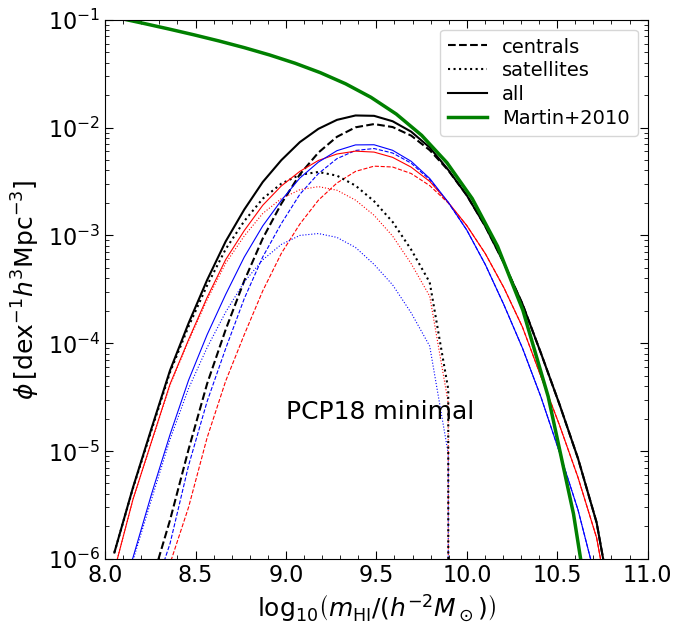}
\includegraphics[width=0.45\textwidth]{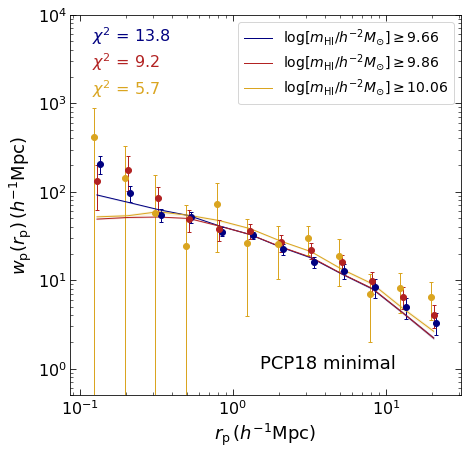}
\caption{\emph{(Left panel:)} Differential \Hi\ mass function of red/blue/all central/satellite/all galaxies in averaged over 6 `minimal PCP18' model mocks using the ${\rm L}300\_{\rm N}1024$ configuration. Colour segregation was based on \eqn{eq:gr-cut} applied to $g-r$ and $M_r$ for each galaxy. For comparison, the solid green curves show the corresponding fit from \citet{martin+10} to \Hi-selected galaxies in the ALFALFA survey. \emph{(Right panel:)} Projected 2pcf in the `minimal PCP18' model (averaged over the same mocks as in the left panel) for three \mHi\ thresholds, compared with corresponding ALFALFA measurements from \citet{guo+17}. The labels indicate the values of $\chi^2$ when comparing each mock result with the corresponding 12 data points from the ALFALFA measurements using the covariance matrices kindly provided by Hong Guo.}
\label{fig:Hi-minimal}
\end{figure*}

Figure~\ref{fig:colmag} shows the joint distributions of $M_r$, $g-r$ and $u-r$ in one mock using the ${\rm L}300\_{\rm N}1024$ box. We clearly see the well-known colour-magnitude and colour-colour bimodality, another sanity check on the mock algorithm.

Turning to somewhat more detailed tests, Figure~\ref{fig:Mrm*func} compares the differential luminosity and stellar mass functions averaged over 6 realisations of the ${\rm L}300\_{\rm N}1024$ box with fitting functions to SDSS measurements from the literature. In each case, for the mock measurements we show results separately for red/blue/all central/satellite/all galaxies. The total luminosity function is in reasonable agreement with the Schechter function fit from \citet{blanton+03}, which is not very surprising since the thresholded luminosity function was used as a constraint in the HOD calibration. 
The total stellar mass function of the mocks also agrees reasonably well with the corresponding fit from \citet{peng+12}. In this case, we also have individual fits for centrals and satellites, which were produced by \citet{peng+12} using the SDSS group catalog of \citet{yang+07}, which similarly agree well with the stellar mass functions of mock centrals and satellites, respectively.
Considering the substantial amount of systematic uncertainty involved in extracting stellar mass functions from data, as well as inherent systematics in the galaxy classification algorithm used to produce the SDSS group catalog, we conclude that the mocks are in good agreement with the data here as well.

\begin{figure*}
\centering
\includegraphics[width=0.45\textwidth]{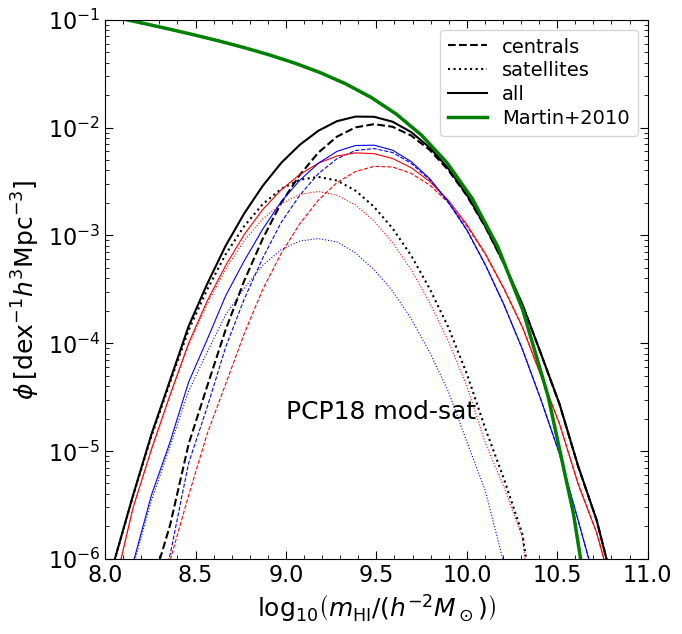}
\includegraphics[width=0.45\textwidth]{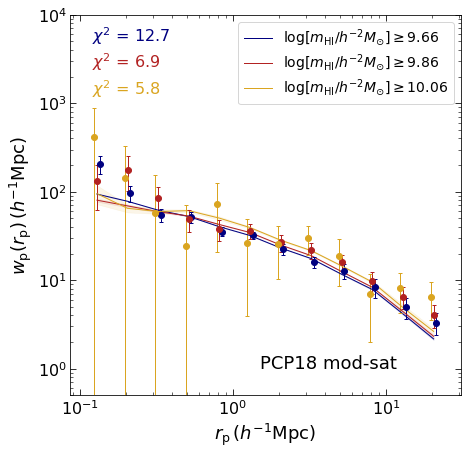}
\caption{Same as figure~\ref{fig:Hi-minimal}, showing results for the `PCP18 mod-sat' model described in section~\ref{subsec:algorithm}, which discards \Hi-selected satellites in parent halos with $m\geq m_{\rm sat,max}$. The value of $m_{\rm sat,max}$ was set to $10^{14.4}\Mh$ by minimising the $\chi^2$ between the mock 2pcf results and corresponding ALFALFA measurements and covariance for the lowest mass threshold shown. We adopt this as our default model for assigning \Hi\ mass to mock galaxies.}
\label{fig:Hi-modsat}
\end{figure*}

\subsection{Neutral hydrogen properties}
\label{subsec:Hi}
\noindent
Figure~\ref{fig:Hi-minimal} compares the \Hi\ mass function and projected 2pcf in the `minimal PCP18' model (see section~\ref{subsec:algorithm}) with measurements in the ALFALFA survey. The \emph{left panel} shows the differential \Hi\ mass function of red/blue/all central/satellite/all galaxies averaged over 6 realisations of the ${\rm L}300\_{\rm N}1024$ box, compared with the fit to \Hi-selected galaxies in the ALFALFA survey by \citet{martin+10}. We see reasonable agreement above the completeness limit of the catalog \citep[set by the luminosity completeness threshold; see section~\ref{subsec:mHI} and the discussion in][]{pcp18}. The \emph{right panel} shows that the 2pcf of mock galaxies with $\log_{10}(\mHi/\Mhsq)>10.0+2\log_{10}(0.678)$ compares slightly worse with the corresponding ALFALFA measurements from \citet{guo+17} than the higher \mHi\ thresholds, which perform well. The 2pcf for \Hi-selected mock galaxies was calculated using \eqn{eq:wprp} with $\pi_{\rm max}=20\Mpch$ to match the \citet{guo+17} measurements.

As mentioned earlier, this slight disagreement is likely due to our use of an updated and improved optical HOD. We therefore explore the modification of the PCP18 model described in section~\ref{subsec:algorithm} and discard \Hi-selected satellites in parent halos with $m\geq m_{\rm sat,max}$. To set the value of the threshold, we attempted to minimise the $\chi^2$ between mocks and data for the threshold $\log_{10}(\mHi/\Mhsq)\geq10.0+2\log_{10}(0.678)$. We found that the $\chi^2$ has a very broad minimum in the vicinity of $m_{\rm sat,max}=10^{14.4}\Mh$, which we use as our default value.
The resulting \Hi\ mass function and projected 2pcf are shown in Figure~\ref{fig:Hi-modsat}; we see a mild improvement in the 2pcf of the lowest threshold, and also some improvement in the higher thresholds. Since this halo thresholded model, which we refer to as `PCP18 mod-sat', allows for the existence of massive satellites while still agreeing with observations, we choose to adopt it as our default model.
For comparison, the number densities in units of $(\Mpch)^{-3}$ in this model for each thresholded sample (in order of increasing threshold \Hi\ mass) in the mocks are, respectively, $\{2.274\pm0.003,0.810\pm0.002,0.203\pm0.0007\}\times10^{-3}$ (with errors estimated using the scatter across 6 realisations), while the corresponding values from Table~1 of \citet{guo+17} are $\{2.68,0.92,0.22\}\times10^{-3}$ (no errors are provided on these values).

We have also explored several `beyond halo mass' modifications of the PCP18 model by changing the criterion used for discarding \Hi-selected satellites. In particular, we considered thresholds on (i) halo mass and concentration jointly (i.e., only allowing satellites in low mass and high concentration halos), (ii) halo concentration alone (only high concentration halos allowed), (iii) large-scale linear halo bias (low bias halos allowed) and (iv) discarding all `red mode' satellites. These are generally inspired by the results of \citet{guo+17}, who found that abundance matching preferentially younger (sub)halos with \Hi-selected galaxies led to good descriptions of ALFALFA clustering. Of these, a joint threshold on halo mass and concentration performs the best, but leads to minimum $\chi^2$ values  nearly identical to those for the `PCP18 mod-sat' model above, at the cost of one additional parameter. We therefore conclude that the ALFALFA data for massive \Hi\ galaxies \emph{do not require `beyond halo mass' effects} in modelling the mass function and projected 2pcf, provided that \Hi\ mass is assigned through an optical scaling relation.

\section{Predictions and extensions}
\label{sec:predict_extend}

\begin{figure*}
\centering
\includegraphics[width=0.475\textwidth,trim=8 10 5 5,clip]{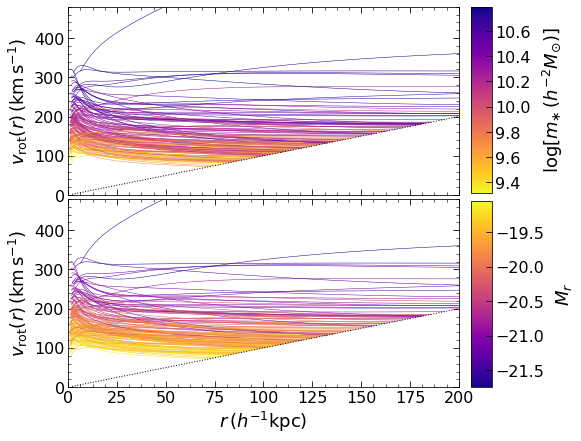}
\hskip 0.05in
\includegraphics[width=0.475\textwidth,trim=2 10 5 5,clip]{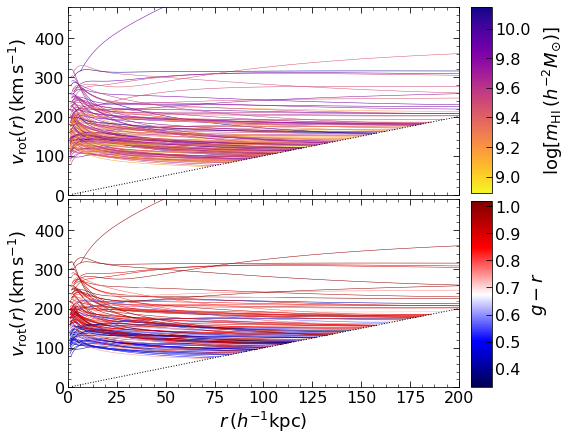}
\caption{Rotation curves of 150 central galaxies containing \Hi\ disks, chosen at random from one mock using the ${\rm L}300\_{\rm N}1024$ configuration. Each curve is coloured by the galaxy's stellar mass $m_\ast$ \emph{(top left)}, luminosity $M_r$ \emph{(bottom left)}, \Hi\ mass $m_{\Hi}$ \emph{(top right)} and colour $g-r$ \emph{(bottom right)}. Since each rotation curve is only generated for $r\leq R_{\rm vir}$, the curves truncate at the dotted line which shows $v(r) = (V_{\rm vir}/R_{\rm vir})\,r$ in each panel. }
\label{fig:rotationcurves}
\end{figure*}

\begin{figure}
\centering
\includegraphics[width=0.45\textwidth,trim=5 5 5 5,clip]{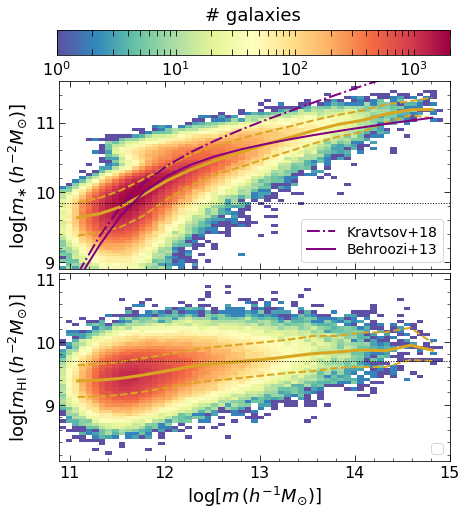}
\caption{Correlation with halo mass. Histograms show the $m_\ast-m$ \emph{(top panel)} and $m_{\Hi}-m$ \emph{(bottom panel)} relation for all central galaxies with $M_r\leq-19$ in one mock using the ${\rm L}300\_{\rm N}1024$ configuration. Solid yellow lines in each panel show the median relation in bins of halo mass, while dashed yellow lines show the corresponding $16^{\rm th}$ and $84^{\rm th}$ percentiles. For calculating these curves, in the \emph{top  panel}, we ignore halos which do not contain a central galaxy and in the \emph{bottom panel}, we further ignore halos whose central does not contain any \Hi\ mass. The solid purple curve in the \emph{top panel} shows the SHAM calibration from \citet{behroozi+13-SHAM}, while the dash-dotted purple curve shows the same relation with parameters taken from \citet{kvm18}, converted to the $m_{\rm 200b}$ mass definition in each case. 
The horizontal dotted lines in each panel indicate the approximate completeness thresholds for $m_\ast$ and $m_{\Hi}$ in the mock.}
\label{fig:mstarmHImhalo}
\end{figure}

Having demonstrated that our mock catalogs reproduce the basic 1-point and 2-point observables associated with galaxy samples selected by optical or \Hi\ properties, in this section we discuss certain predictions of our mocks, along with a few possible extensions.

\subsection{Rotation curves and baryon mass-halo mass relations}
\label{subsec:galDMconnection}

Figure~\ref{fig:rotationcurves} shows the rotation curves of 150 randomly chosen mock central galaxies containing \Hi\ disks, with the curves in each panel being coloured by one of $m_\ast$, $M_r$, $m_{\Hi}$ or $g-r$. As noted earlier, these are generally flat but show considerable diversity, in qualitative agreement with observed rotation curves \citep{pss96,mrdb01}. We defer a more quantitative comparison with observations to future work.

\begin{figure*}
\centering
\includegraphics[width=0.48\textwidth]{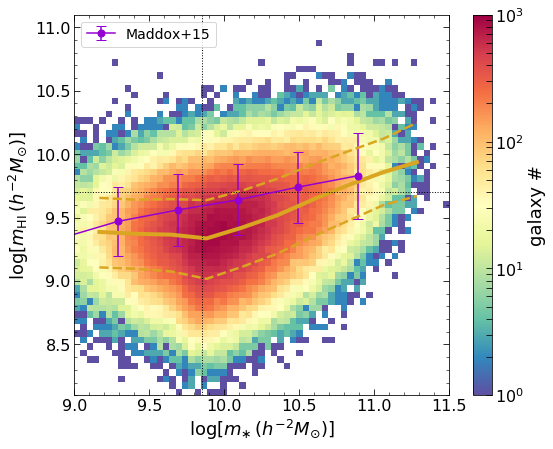}
\includegraphics[width=0.41\textwidth]{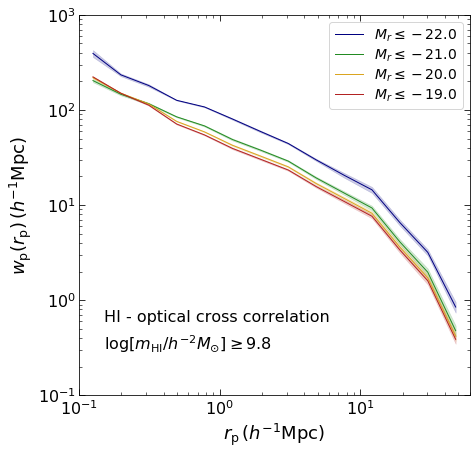}
\caption{Predictions of optical-\Hi\ correlations not used in constraining the galaxy-dark matter connection in our mocks. \emph{(Left panel:)} Joint distribution (coloured histogram) of $m_{\Hi}$ and $m_\ast$ for galaxies containing \Hi\ in one mock using the ${\rm L}300\_{\rm N}1024$ box. Solid yellow line indicates the median $m_{\Hi}$ in bins of $m_\ast$, while dashed yellow lines indicate the $16^{\rm th}$ and $84^{\rm th}$ percentiles. Vertical and horizontal dotted lines indicate the completeness limits of the mock in $m_\ast$ and $m_{\Hi}$, respectively (see Figures~\ref{fig:Mrm*func} and~\ref{fig:Hi-modsat}). For comparison, the purple symbols with error bars  show the relation calibrated by  \citet[][see their Table~1]{maddox+15} using a cross-matched sample of \Hi-selected galaxies from the ALFALFA and SDSS surveys. \emph{(Right panel:)} 2-point projected cross-correlation function between galaxy samples selected by optical luminosity thresholds (indicated by colours) and a sample selected by \Hi\ mass threshold (indicated in the label) using 3 realisations of the ${\rm L}300\_{\rm N}1024$ configuration. The samples lie in the same volume and therefore overlap in membership, but are not explicitly cross-matched during their selection.}
\label{fig:Hi-predict}
\end{figure*}

Our mocks also predict the relations between group halo mass and stellar mass ($m_\ast-m$) as well as \Hi\ mass ($m_{\Hi}-m$). Figure~\ref{fig:mstarmHImhalo} shows the $m_\ast-m$ relation \emph{(top panel)} and the $m_{\Hi}-m$ relation \emph{(bottom panel)} for all central galaxies with $M_r\leq-19$ in one mock using the ${\rm L}300\_{\rm N}1024$ configuration. The \emph{top panel} is essentially the same as Figure~A4 of \citet{pkhp15}, except that we have used an updated HOD. Similarly, the \emph{bottom panel} can be compared with Figure~B1 of PCP18, who showed the median $m_{\Hi}-m$ relation using their analytical halo model. Both sets of results are consistent with these earlier works, showing a steep $m_\ast-m$ relation but a much shallower $m_{\Hi}-m$ relation. The latter feature also emphasizes the need for caution when painting \Hi\ directly into halos: the weak correlation between \Hi\ mass and halo mass can amplify systematic errors in any calibration.

For comparison, the purple curves in the \emph{top panel} show the median $m_\ast-m$ relations calibrated using SHAM by \citet[][solid]{behroozi+13-SHAM} and \citet[][dash-dotted]{kvm18}, converted in each case to the $m_{\rm 200b}$ mass definition appropriate for this work. We see that, for stellar masses above the completeness limit of our mocks (horizontal dotted line), the mock result is closer to the \citet{behroozi+13-SHAM} relation for $m\lesssim10^{13.5}\Mh$ and lies between the two SHAM calibrations at larger halo masses. 
A similar comparison with the literature for the $m_{\Hi}-m$ relation is complicated by the fact that different authors have used different conventions for defining this relation \citep[e.g.,][]{pra17,guo+17}. We have checked that our results are qualitatively similar to these calibrations, leaving a more detailed analysis to future work.

\subsection{HI-optical cross-correlations}
A primary strength of our mock algorithm is its ability to paint realistic optical and \Hi\ properties \emph{in the same galaxies}. This means that we can go beyond previous studies and predict or forecast expectations for the \emph{joint} distribution of, say, stellar and \Hi\ mass in low-redshift galaxies, along with the corresponding spatial correlations.

Figure~\ref{fig:Hi-predict} shows the $m_{\Hi}-m_\ast$ relation \emph{(left panel)} and the spatial cross-correlation function between galaxy samples selected by luminosity and \Hi\ mass \emph{(right panel)} in our mocks. These are genuine predictions of our algorithm; comparing these with corresponding measurements forms a test of the various underlying assumptions. Indeed, as already noted by PCP18, we see in the left panel that the predicted $m_{\Hi}-m_\ast$ relation is in reasonable agreement with the results  of  \citet[][purple points with errors]{maddox+15}, although the median trend in the mocks is slightly lower than in the data. Note, however, that our mocks are only complete above  the $m_\ast$ and $m_{\Hi}$ thresholds indicated by the vertical and horizontal dotted lines, respectively. As such, a robust comparison with observations is not possible in the mass range we can explore.

To date, the only measurements of cross-correlations similar to those in the \emph{right panel} are by \citet{papastergis+13}, who studied the projected cross-2pcf between \Hi-selected galaxies in ALFALFA and colour-selected galaxies in SDSS (see their Figures~17 and~18). As pointed out by \citet{guo+17}, however, the weights used by \citet{papastergis+13} in their 2pcf measurements did not accurately account for sample variance effects, which are substantial in the small volume ($z\lesssim0.05$) probed by the ALFALFA survey. E.g., \citet{guo+17} reported a significant $m_{\Hi}$-dependence of clustering, which was not detected by \citet{papastergis+13}. 

It will therefore be very interesting to confront our mock catalogs with more robust cross-2pcf measurements.
For example, the choices controlling \Hi-satellites in our model (namely, the value of the threshold halo mass $m_{\rm sat,max}$) affect the shape of the cross-correlation at small separations between bright optical galaxies and \Hi-selected galaxies and can therefore be tested by such observations.

\subsection{Possible extensions}
Although the mocks we have presented here provide fairly realistic descriptions of the distribution of optical and \Hi\ properties of local Universe galaxies, they contain several ingredients which can be potentially improved upon or extended. We list some of these here.

\subsubsection{Assembly bias}
We noted in section~\ref{subsec:algorithm} that our assignment of halo concentrations $c_{\rm 200b}$ preserves spatial correlations of $c_{\rm 200b}$ at fixed halo mass (also called assembly bias) under the assumption that these correlations are accurately tracked even by poorly resolved halos. Recently, \citet{rps21} have shown that this assumption fails at worse than $\sim20\%$ for halos resolved with $\lesssim150$ particles. Instead, they showed that the spatial correlations of the local tidal anisotropy $\alpha$ are accurately preserved even for halos with as few as 30 particles. Using their technique of sampling a  conditional distribution $p(c_{\rm 200b}|m,\alpha)$, therefore, will be a promising extension of our algorithm that would endow mocks built on low-resolution $N$-body simulations with accurate representations of halo assembly bias. 
%
Another interesting extension, also easy to include in our mocks, would be an environment-dependent modulation of the HOD itself, as discussed by \citet{xzc21}. Combined with the conditional sampling of halo properties, this would lead to full flexibility in modelling galaxy assembly bias.

\subsubsection{Stellar disk-bulge decomposition}
We also noted above that our treatment of the stellar spatial profile is, strictly speaking, inconsistent because it assumes a spherically symmetric distribution of stars rather than, say, an axially symmetric disk. 
Since our primary intention is to produce a rotation curve for each galaxy, 
the difference between a spherical bulge and an axial disk can potentially be a large effect 
\citep{binney-tremaine-GalDyn}. 
We intend to explore this further in a forthcoming work, by simultaneously modelling a  stellar disk and bulge using realistic bulge-to-disk mass ratios \citep[e.g.][]{bernardi+14}, along with their correlations with galaxy colours and the presence of an \Hi\ disk. 

\subsubsection{Gas fractions}
In this work, we used the expression \eqref{eq:fbgas} for the bound gas fraction $f_{\rm bgas}$, with parameters adopted from ST15, for modelling all central galaxies with $M_r\leq-19$. Strictly speaking, this relation holds for the central galaxies of halos with $m_{\rm vir}\gtrsim10^{13}\Mh$, since it is calibrated using X-ray observations of galaxy clusters. Our choice therefore corresponds to an extrapolation of this relation into an unobserved regime of halo mass.
Since the resulting value of $f_{\rm bgas}$ for each central is typically substantially smaller than its $f_{\rm egas}$ (which is set by baryonic mass conservation in this work), we do not expect this extrapolation to lead to any significant systematic error for any of  the statistics explored in this paper. Improvements to this model could potentially explore using observations of the circum-galactic medium to first constrain $f_{\rm egas}$.

\subsubsection{Predictions at higher redshift}
All of our results have been restricted to the local Universe, a consequence of using clustering constraints from the low-redshift ($z\lesssim0.1$) surveys SDSS and ALFALFA. 
It will be interesting to extend our results to the redshift range $0.5\lesssim z\lesssim 1$,
which is interesting for both astrophysics and cosmology, and is the target of several completed, ongoing and upcoming galaxy surveys.
In future work, we will explore whether an extension of our low-redshift algorithm, augmented by simplified galaxy evolution models \citep[e.g.,][]{lilly+13}, can be used to make robust predictions at these higher redshifts.

\section{Conclusion}
\label{sec:conclude}
The ability to realistically reproduce, in a simulated universe, the properties and spatial distribution of galaxies observed in the actual Universe, opens the door to addressing a number of interesting astrophysical and cosmological questions. 
We have presented an updated algorithm that produces catalogs of mock galaxies in simulated halos at $z\approx0$, realistically endowed with a variety of properties including $r$-band luminosities, $g-r$ and $u-r$ colours, stellar masses $m_\ast$, neutral hydrogen (\Hi) masses $m_{\Hi}$, as well as (for central galaxies) the spatial distribution of gas and stars, leading to realistic rotation curves. Our mock galaxies additionally inherit a number of environmental properties from their host dark matter halos, including the halo-centric overdensity, tidal anisotropy and large-scale halo bias.

Our algorithm, which relies on an HOD which assigns galaxy properties based on halo mass $m$ alone, can optionally include effects such as galactic conformity and colour-dependent galaxy assembly bias, and is easily extendable to include effects such as environment-dependent modulations of the HOD. By construction, the basic mocks we presented here reproduce the luminosity function, colour-luminosity relation and the luminosity- and colour-dependent 2-point clustering of optically selected SDSS galaxies with $M_r\leq-19$, as well as the \Hi\ mass function and \Hi-dependent 2-point clustering of \Hi-selected ALFALFA galaxies with $m_{\Hi}\gtrsim10^{9.7}\Mhsq$. The mocks then  reproduce the SDSS stellar mass function and the SDSS-ALFALFA $m_\ast-m_{\Hi}$ relation reasonably well (these were not used when constraining the parameters of the algorithm), while predicting the spatial 2-point cross-correlation function of low-redshift optical and \Hi\ galaxies (which has not yet been robustly measured; Figure~\ref{fig:Hi-predict}), and the $m_\ast-m$ and $m_{\Hi}-m$ relations (Figure~\ref{fig:mstarmHImhalo}). The calibrations we used lead to volume-completeness thresholds of $M_r\leq-19$, $m_\ast\gtrsim10^{9.85}\Mhsq$ and $m_{\Hi}\gtrsim10^{9.7}\Mhsq$, thus representing the population of massive galaxies in the low-redshift Universe.
Our algorithm represents a consolidation of the results of \citet{pkhp15}, \citet{pcp18} and \citet{ppp19}.

Our mocks are potentially useful for a number of  applications, some of which we list here. 
\begin{itemize}
    \item In their study of environment-dependent clustering in SDSS, \citet{phs18b} noted some small ($\sim20\%$) but significant differences between their `mass-only HOD' mocks and SDSS galaxies in the most anisotropic tidal environments. These differences were ultimately inconclusive due to the comparable level of systematic uncertainties associated with the HOD calibration used by \citet{phs18b}. The mocks we have presented, which are based on updated calibrations, largely mitigate many of these uncertainties. It will therefore be interesting to revisit this analysis to assess the level of beyond-mass effects induced by the tidal environment in the SDSS field \citep[see also][]{azpm19}. 
    \item As we noted earlier, our algorithm successfully describes \Hi-dependent clustering at the massive end in ALFALFA without the need of assembly bias, unlike earlier studies \citep[e.g.][]{guo+17}. Our mocks can therefore serve as useful null tests for galaxy assembly bias using interesting new combinations of observables, such as the large-scale bias of galaxies split by \Hi\ mass in bins of stellar mass. 
    \item Galactic conformity, a putative non-local connection between the satellites and central galaxy of the same halo, continues to pose a  puzzle for galaxy formation models \citep{weinmann+06,hbv16}. Our mocks can be used to explore new tests of this phenomenon, e.g., the potential dependence of galactic conformity on tidal environment \citep[which is otherwise an excellent indicator of halo assembly bias, see][]{rphs19}.
    \item Our mocks can predict the \Hi\ mass function of (massive) galaxies selected by optical luminosity or colour. The corresponding measurements have only recently become available \citep{dkd20,dk21}, and will be very useful for testing our basic assumptions regarding the connection between optical properties and \Hi.
    \item The rotation curves of our mock central galaxies, along with their \Hi\ disks when present, can be used to model the observed 21cm velocity profiles of \Hi-selected galaxies. The distribution of the widths of these profiles has been measured in the ALFALFA survey \citep{papastergis+11,moorman+14}  and constitutes an exciting and hitherto unexplored new probe of the small-scale distribution of baryonic matter. 
    \item The `radial acceleration relation' (RAR) between the acceleration profiles due to dark matter and baryons in disk galaxies \citep{mls16,lmsp17} has emerged as an intriguing new probe of gravitational theories at galactic length scales. The rotation curves and mass profiles of baryonic and dark matter in our mock central galaxies enable an exploration of the nature of the RAR for large samples of galaxies in the CDM+baryons framework \citep{ps21}.
\end{itemize}

\section*{Acknowledgments}
We thank R. Srianand for his collaboration and for many valuable discussions and suggestions which improved the presentation of this  paper. AP also  thanks Nishikanta Khandai and Kandaswamy Subramanian for insightful discussions. 
The research of AP is supported by the Associateship Scheme of ICTP, Trieste and the Ramanujan Fellowship awarded by the Department of Science and Technology, Government of India. 
TRC acknowledges support of the Department of Atomic Energy, Government of India, under project no.~12-R\&D-TFR-5.02-0700 and the Associateship Scheme of ICTP. 
This work made extensive use of the open source computing packages NumPy \citep{vanderwalt-numpy},\footnote{\url{http://www.numpy.org}} SciPy \citep{scipy},\footnote{\url{http://www.scipy.org}} Matplotlib \citep{hunter07_matplotlib},\footnote{\url{https://matplotlib.org/}} Jupyter Notebook\footnote{\url{https://jupyter.org}} and the plotting software Veusz.\footnote{\url{https://veusz.github.io/}}
We gratefully acknowledge the use of high performance computing facilities at IUCAA, Pune.\footnote{\url{http://hpc.iucaa.in}}

\section*{Data availability}
The mock catalogs generated by our algorithm will be shared upon reasonable request to the authors.

\bibliography{masterRef,additionalRef}

\appendix

\section{Quasi-adiabatic relaxation}
\label{app:rdm}
Here we describe the technique for calculating the response of the dark matter profile to presence of baryonic matter through an approximate conservation of angular momentum (\citealp{bw84}; \citealp{bffp86}; \citealp{gkkn04}; \citealp{abadi+10}; \citealp{teyssier+11}; ST15). The discussion below follows section 2.3 of ST15 (see their equations 2.15-2.17). 

The basic equation describing this quasi-adiabatic relaxation gives the final radius $r$ of a spherical dark matter element in terms of its initial radius  $r_{\rm in}$,
\be
\frac{r}{r_{\rm in}} = 1 + q_{\rm rdm}\left(\frac{m_{\rm nfw}(<r_{\rm in})}{m_{\rm tot}(<r)} - 1 \right)\,.
\label{eq:qar-def}
\ee
Here, $m_{\rm tot}(<r) = m_{\rm bary}(<r) + m_{\rm rdm}(<r)$ is the total mass contained inside the final radius $r$, with $m_{\rm bary}(<r) = 4\pi\int_0^r\der r^\prime\,r^{\prime2}\,f_{\rm bary}\,\rho_{\rm bary}(r^\prime)$ being the baryonic component and $m_{\rm rdm}(<r)$ being the final, relaxed dark matter mass profile which satisfies
\be
m_{\rm rdm}(<r) = f_{\rm rdm}\,m_{\rm nfw}(<r_{\rm in})\,,
\label{eq:m_rdm}
\ee
and $m_{\rm nfw}(<r_{\rm in}) = 4\pi\int_0^{r_{\rm in}}\der r^\prime\,r^{\prime2}\,\rho_{\rm nfw}(r^\prime)$ is the dark matter mass inside the \emph{initial} radius as per the original, normalised NFW profile. We remind the reader that all the density profiles are normalised so as to enclose the entire mass $m_{\rm vir}$  inside $r=R_{\rm vir}$.

The quantity $q_{\rm rdm}$ is a parameter controlling the level of angular momentum conservation. From  \eqn{eq:qar-def}, we see that $q_{\rm rdm}=1$ corresponds to perfect conservation, since $r\,m(<r)$ is an adiabatic invariant in this case.
On the other hand, $q_{\rm rdm}=0$ corresponds to no baryonic backreaction. In this work, we follow ST15 and set $q_{\rm rdm}=0.68$, which has been found to accurately describe the cumulative  effects of baryonic backreaction effects both in the inner and outer regions of simulated halos, accounting for the fact that the formation of the central galaxy is not instantaneous. The effect of varying $q_{\rm rdm}$ on rotation curves and related statistics will be the focus of a future study.

Defining the ratio
\be
\xi \equiv r/r_{\rm in}\,,
\label{eq:xi-def}
\ee
\eqn{eq:qar-def} can be re-written as
\begin{align}
\Cal{L}(\xi|r) &\equiv \xi - 1 + q_{\rm rdm} - \frac{q_{\rm rdm}}{f_{\rm rdm}}\left[1 + \frac{m_{\rm bary}(<r)}{f_{\rm rdm}m_{\rm nfw}(<r/\xi)}\right]^{-1}\notag\\ 
&= 0\,,
\label{eq:qar-alt}
\end{align}
which is conducive to an iterative solution. We employ Newton's method using an analytical expression for the derivative $\Cal{L}^\prime(\xi|r)=\p\Cal{L}(\xi|r)/\p\xi$ at fixed $r$ and writing the estimate at the $n^{\rm th}$ iteration as
\be
\xi^{(n)} = \xi^{(n-1)} - \frac{\Cal{L}(\xi^{(n-1)}|r)}{\Cal{L}^\prime(\xi^{(n-1)}|r)}\,.
\label{eq:newton}
\ee
For practically all baryonic configurations and values of $r\in(10^{-3},1)\times R_{\rm vir}$, and for all values $0\leq q_{\rm rdm}\leq 1$, convergence is achieved with a  relative tolerance of $10^{-5}$ in $\lesssim8$ iterations using \eqn{eq:newton}. (In contrast, a simple iteration applied directly to equation~\ref{eq:qar-alt} typically requires several tens to hundreds of iterations for $q_{\rm rdm}\lesssim0.8$, while the inner regions of the halo do not converge for $q_{\rm rdm}\gtrsim0.9$.)

Knowing the ratio $\xi$ at any $r$ then gives the mass of relaxed dark matter enclosed in radius $r$ using \eqn{eq:m_rdm} setting $r_{\rm in}=r/\xi$ on the right hand side.
By construction, $\xi=1$ at $r=R_{\rm vir}$, so that $m_{\rm rdm}(<R_{\rm vir}) = f_{\rm rdm}\,m_{\rm vir}$, as it  should be. The value of $m_{\rm rdm}(<r)$ at any $r$ is sufficient for calculating the rotation curve $v_{\rm rot}(r)$ using \eqn{eq:vrot}. For the differential profile shown in Figure~\ref{fig:baryonification}, we must differentiate \eqn{eq:m_rdm} with respect to $r$. The (somewhat cumbersome) result can be written analytically entirely in terms of $\xi$ and $r$; we omit it for brevity.

\label{lastpage}

\end{document}